\documentclass[%
 reprint,
 amsmath,amssymb,
 aps,
 prd,
]{revtex4-2}

\usepackage{orcidlink} 
\usepackage{graphicx}
\usepackage{xcolor}
\usepackage{xspace}
\usepackage{bm}
\usepackage{hyperref}

\newcommand{\colhead}[1]{\multicolumn{1}{c}{#1}}

\renewcommand{\selectlanguage}[1]{}

\usepackage{xspace}
\usepackage{bm}
\usepackage{amsmath}
\usepackage{hyperref}

\newcommand{\Tobs}{T_\mathrm{obs}}

\renewcommand{\d}{\mathrm{d}}

\newcommand{\Msun}{\ensuremath{M_\odot}\xspace}

\newcommand{\forb}{\ensuremath{f_\mathrm{orb}}\xspace}

\newcommand{\rhothr}{\rho_\mathrm{thr}}
\newcommand{\rhoopt}{\rho_\mathrm{opt}}

\newcommand{\Tvis}{T_\mathrm{vis}}

\newcommand{\VT}{\mathrm{VT}}

\newcommand{\pdet}{p_\mathrm{det}(\bm\theta, z, \tau)}

\newcommand{\rate}{{\mathcal{R}}}

\newcommand{\Ndet}{N_\mathrm{det}}

\newcommand{\inspbon}{\zeta(\bm\theta)}
\newcommand{\gpcyr}{\ensuremath{\mathrm{Gpc}^3\,\mathrm{yr}}}

\newcommand{\mwd}{m_\mathrm{wd}}
\newcommand{\Rwd}{R_\mathrm{wd}}

\newcommand{\LIMRI}{CO-IMBH\xspace}
\newcommand{\KerrEq}{\texttt{KerrEq}\xspace}
\newcommand{\AAK}{\texttt{AAK-PN5}\xspace}
\newcommand{\FEW}{\texttt{FEW}\xspace}

\newcommand{\ib}{{no-plunge bonus}\xspace}

\begin{document}

\title{Long before plunge: Uncovering intermediate-mass black holes through their inspiralling compact object companions with LISA}

\author{Jean Maurice Meus
\orcidlink{0009-0005-9338-7623}}
\email{jmm974@student.bham.ac.uk}

\author{Christian Chapman-Bird
\orcidlink{0000-0002-2728-9612}}
\email{c.chapman-bird@bham.ac.uk}

\author{Alberto Vecchio
\orcidlink{0000-0002-6254-1617}}
\email{a.vecchio@bham.ac.uk}
\affiliation{Institute for Gravitational Wave Astronomy \& School of Physics and Astronomy, University of Birmingham, Birmingham, United Kingdom}

\date{\today}

\begin{abstract}

The Laser Interferometer Space Antenna (LISA) will provide novel capability to discover and study intermediate-mass ($\sim 10^2-10^5\,\Msun$) black holes at the centres of dwarf galaxies and in dense stellar environments. 
We systematically map out the LISA-detectable parameter space of binaries consisting of a low-mass compact object spiralling into an intermediate-mass black hole (mass ratio $\sim 10^{-5} - 10^{-2}$) on equatorial orbits with generic eccentricity.
The depth of the LISA survey reaches a redshift $\approx 5$ ($0.1$) for binaries with a primary black hole mass $m_1 \approx 10^4\,\Msun$ ($ 10^3\,\Msun$). 
If the orbital companion is a white dwarf, which could be disrupted prior to plunge, LISA's survey depth is well matched to (and, in many cases, exceeds) the one of electro-magnetic surveys. 
Although the maximum distance at which a source can be detected depends weakly on the orbital eccentricity, the latter strongly affects the portion of the binary inspiral over which a source is detectable.
The instrument's sensitivity time-volume varies by many orders of magnitude depending on the binary's component masses: it reaches $\sim 1 - 10^3\,\gpcyr$ ($\sim 10^{-2} - 10^3\,\gpcyr$) for the more massive systems with $m_1 = 10^5\,\Msun$ ($ 10^4\,\Msun$) and decreases to $\sim 10^{-4} - 1\,\gpcyr$ for lighter primary black holes with $m_1 = 10^3\,\Msun$. 
Importantly, from tens of percent to the overwhelming majority of the total sensitivity time-volume (and therefore number of potential detections) comes from sources that do not plunge during the yr-long mission lifetime and are observed at an earlier stage of their inspiral. 
Despite the poorly known merger rates of this class of sources, our results show that LISA could identify from a handful to $\mathcal{O}(10^3)$ intermediate-mass black holes providing a new observational avenue for studying these elusive objects and the environments in which they reside.

\end{abstract}

\maketitle

\section{Introduction}
\label{sec:intro}

It is well established that astrophysical processes form black holes (BHs) over a wide mass spectrum. \textit{Stellar-mass} ($\sim 1-100\,\Msun$) BHs are known in the Milky Way through monitoring of X-ray binaries~\citep{Corral-Santana+2016A&A}, astrometry~\citep{GAIA-BH1, GAIA-BH2, GAIA-BH3, Wang+2024}, possibly radio-pulsars~\citep{Barr+2024} and at increasingly higher redshift via ground-based gravitational-wave (GW) observations~\citep{GW150914, GWTC5-catalogue, GWTC5-pop}.
Evidence for \textit{(super-)massive} ($\sim 10^{5}-10^{10}\,\Msun$) BHs has been continuously growing with striking examples in Sagittarius A$^*$~\citep{EckartGenzel1997, Schoedel+2002, Ghez+2008, Gillessen+2009, EHT-SgrAshadow2022, GRAVITY2022, ElDayem+2026} and M87~\citep{Gebhardt+2011, Walsh+2013, EHT-M87shadow2019}, and with candidates from optical spectroscopic surveys~\citep{WuShen2022} and tidal disruption events (TDEs)~\cite{Hills1975, Rees1988, Wevers+2017, Grotova+2025, Ramsden+2026}.

Observational evidence for``intermediate-mass" BHs ($\sim 10^{2}-10^{5}\,\Msun$; IMBHs) remains, however, elusive, see \textit{e.g.} Refs.~\cite{Mezcua2017, Greene+2020} and references therein. Candidates are known from dynamical measurements of stars or accreting gas in dense environments such as globular clusters and star clusters, broad-line AGNs in dwarf galaxies, compact radio cores and possibly micro-lensing observations. Time-domain variability~\cite{BurkeNatarajan2026} and reverberation mapping~\cite{Guo+2026} surveys may offer opportunities to discover these elusive objects in the future. 

It is of central importance for astrophysics to establish whether the current ``observational desert" of BHs at intermediate masses is simply due to the observational challenges of unveiling these objects with present instruments, or rather reflects an intrinsic paucity of BHs.   
In fact, understanding the properties of IMBHs and those of the environments in which they reside is essential to answer key questions such as how BHs form and evolve, the initial seeding of BHs that eventually evolve into the present-day population of super-massive BHs, the dynamics and evolution of stars and dense environments (globular clusters and nuclear star clusters), and feedback mechanisms in (dwarf) galaxies, see Refs.~\cite{Mezcua2017, Greene+2020} and references therein

In this paper, we address the question of how GW observations with the Laser Interferometer Space Antenna (LISA)~\citep{LISA, LISA_DSR} may contribute to discover evidence for BHs with mass $\sim 10^{2}-10^{5}\,\Msun$. Although we focus on LISA, our conclusions are also relevant to other planned space-based detectors, such as Taiji~\citep{Taiji2021} and TianQin~\citep{TianQinReview2024}.

IMBHs have been recognised for a long time as potential GW sources~\citep{Miller2002, Mandel+2008}. 
Some studies have already considered how LISA could provide observational evidence for IMBHs, but they have been primarily concerned with binary systems of comparable or moderate (\textit{i.e.} $\gtrsim 10^{-2}$) mass ratio, \textit{e.g.}~\citep{Amaro-Seoane2018, Jani+2020}.
[Our convention is that the mass ratio is $m_2/m_1 \le 1$, where $m_1 \ge m_2$ are the binary's primary and secondary mass, respectively.] 

Here we focus on the less explored source class where a low-mass compact object (CO) inspirals onto  an IMBH, and we shall call these binaries CO-IMBHs.
We set (with some degree of arbitrarity) the mass range covered by these sources to be a primary binary BH with mass $\sim 10^2 - 10^5\,\Msun$ with an orbital CO with a mass $0.1 - 10^3\,\Msun$,  subject to the constraint that the mass-ratio is the range $\approx 10^{-5} - 10^{-2}$,
\textit{i.e.} such systems belongs to what is commonly referred to as \textit{intermediate mass-ratio inspirals}. 
These binaries can form in many different environments, and through a variety of mechanisms~\cite{Fragione+2018, ArcaSeddaCapuzzoDolcetta2019, Rasskazov+2020, ArcaSedda+2021, FragioneLoeb2023}. 
The merger rate, $\rate$, and its dependence on the physical parameters of these systems, remain highly uncertain due to the limited understanding of the astrophysical processes surrounding the formation and evolution of these binaries. Estimates range from $\rate \sim 10^{-3} - 10^2\,\mathrm{Gpc}^{-3}\,\mathrm{yr}^{-1}$, the uncertainty further highlighting the importance of observational efforts targeting these systems and the science return from potential discoveries.

An additional element of interest for \LIMRI{}s, as well as extreme mass-ratio inspirals~\citep[EMRIs;][]{BarakCutkler2004, Babak+2017, Gair+2017} is their possible connection with several electro-magnetic transient phenomena. 
Several models are being considered to explain the observations of quasi-periodic-eruptions (QPEs)~\cite{Miniutti+2019, Giustini+2020, Arcodia+2021, Arcodia+2024, Nicholl+2024, Chakraborty+2025, Hernandez-Garcia+2025}, and a possible scenario that is gaining significant attention is the repeated collisions of a compact object ``punching through" material accreting onto a BH to produce the observed X-ray flares~\citep{King2020, King2022, King2023, Franchini+2023, LinialMetzger2023, LinialMetzger2024, Zhou+2024, Yao+2025, Zhou+2025a, Zhou+2025b, Yao2Quataert2026}. 
In the context of tidal disruption events (TDEs), if the compact object orbiting an IMBH is a white-dwarf (WD) it may be tidally disrupted by the IMBH~\citep{Rosswog+2009, MacLeod+2014, MacLeod+2016} and this process could account for some of the transients that have been observed to-date.  
Some unusual high-energy transients recently observed are also amenable of being interpreted as a WD disrupted by an IMBH~\citep{King2020, Eyles-Ferris+2026}. All these astrophysical scenarios could provide fertile ground for multi-messenger observations~\cite{Sesana+2008}. 
While this specific aspect is beyond the scope of this paper and will not be considered here, it is a further example of the multi-faceted role of \LIMRI{}s in future observational campaigns.

\begin{figure}[htb!]
    \centering
    \includegraphics[width=0.99\linewidth]{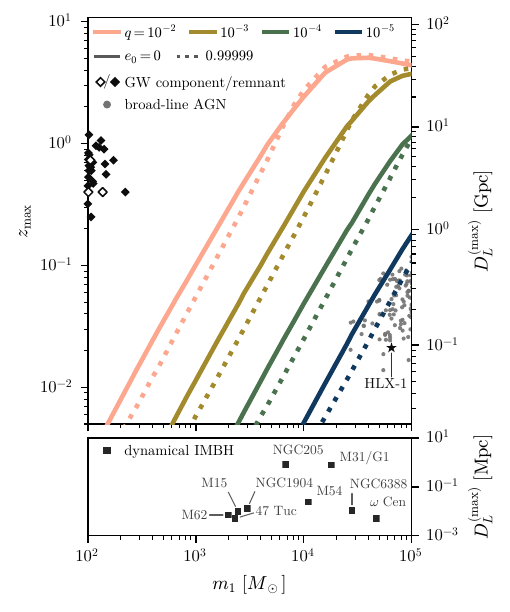}
    \caption{The sky- and polarisation-averaged maximum redshift ($z_\mathrm{max}$) and corresponding luminosity distance ($D^{(\mathrm{max})}_L$), see Eq.~\eqref{eq:horiz}, at which LISA can detect \LIMRI{}s as a function of the (source-frame) primary mass $m_1$, for mass ratios $q = 10^{-2}, 10^{-3}, 10^{-4}, 10^{-5}$ (pink, golden, green and blue, respectively) and eccentricities $e_0 = 0$ (solid lines) and $e_0 = 0.99999$ (dotted lines) at time-to-plunge $\tau_0 = 10^3\,\mathrm{yr}$. The observation time is $\Tobs = 4\,\mathrm{yr}$ and the detection threshold is set to $\rhothr = 15$. Markers show current IMBH candidates in the compilation of Refs.~\cite{BurkeNatarajan2026,burke_bh_mass_spectrum_repo}: broad-line AGN IMBH candidates with single-epoch masses (extended with  Ref. \cite{Guo+2026}); dynamically measured IMBHs (all below the lowest horizon curve); HLX-1; and all GW events with astrophysical probability $> 0.5$, component and remnant masses, included in GWTC-5.0~\citep{GWTC5-catalogue} via GWOSC~\citep{GWOSC}.}
    \label{fig:horizon_vs_M}
\end{figure}

\begin{figure*}[htb!]
    \centering
    \includegraphics[width=0.99\linewidth]{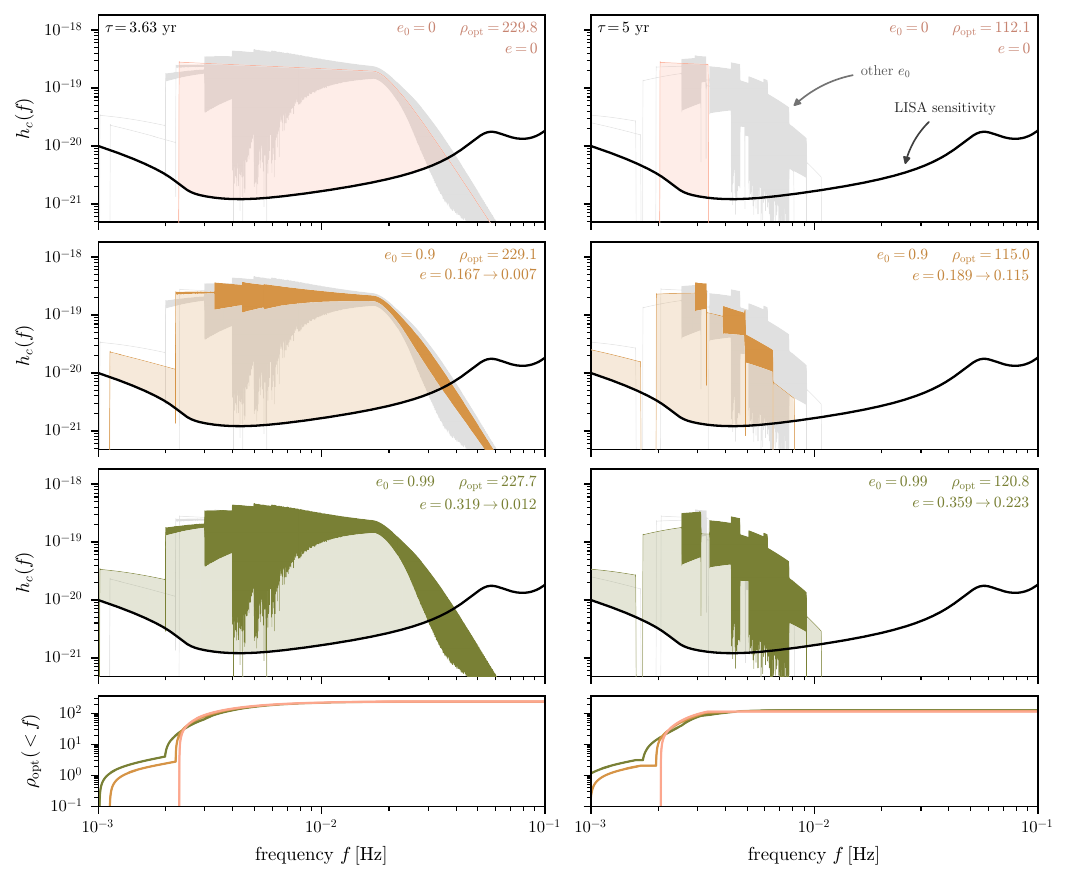}
    \caption{The characteristic strain produced by a non-spinning \LIMRI{} of mass $m_1 = 10^4\,\Msun$, $m_2 = 10^2\,\Msun$ at 
    $z = 0.11$
     observed for $\Tobs = 4\,\mathrm{yr}$ at two different stages of its lifetime. Panels on the left (right) correspond to the observing period starting when the binary's remaining lifetime is $\tau = 3.63\,\mathrm{yr}$ and correspondingly $\tau^{(\mathrm{obs})} = 4.02 \,\mathrm{yr} \simeq \Tobs$ ($\tau = 5\,\mathrm{yr}$ and $\tau^{(\mathrm{obs})} = 5.55\,\mathrm{yr}$), as indicated in the legend. Each row refers to a different eccentricity of the system -- $e_0 = 0$ (pink), $0.9$ (orange) and $0.99$ (green) -- at a reference (source-frame) time-to-plunge $\tau_0 = 10^3\,\mathrm{yr}$. In order to visually emphasise the difference of the signal depending on the eccentricity, the pale grey band in each panel is the envelope of $\tilde h_c(f)$ for the other two eccentricities. Each panel is annotated with $e_0$ and $\rhoopt$, Eq.~\eqref{eq:snr_opt}, as well as the range of eccentricity swept from the beginning to the end of the observation. The solid black curve is the LISA sensitivity. The bottom row shows the SNR accumulated as a function of frequency, $\rhoopt(<f)$, for the three eccentricities.}
    \label{fig:hf_c}
\end{figure*}

The main motivation of this study is to systematically map out the LISA detectable parameter space of \LIMRI{}s, and therefore inform the development of suitable analysis approaches and underpinning GW waveform models able to maximise the detection opportunities of what could turn out to be a unique observational sample. 
In fact, LISA could provide a wider parameter space for discoveries, reaching significantly deeper into the universe, and covering a broader range of IMBH masses than current and future telescopes, as we show in Fig.~\ref{fig:horizon_vs_M}. 
\LIMRI{}s are long-lived sources in the LISA band and, as we exemplify in Fig.~\ref{fig:hf_c}, their emitted gravitational radiation is complex as it is produced by binaries that may in some cases retain some residual eccentricity in the detection band. 
As a consequence, the regions of parameter space most accessible to the instrument and most promising for detection are not immediately obvious.

To address LISA's discovery potential for \LIMRI{}s, we consider two main figures of merit that simply depend on the instrument performance (and are agnostic to the \LIMRI's merger rate) that we dub (i) the \textit{sensitivity time-volume}, $\VT(\bm\theta)$, and (ii) the \textit{\ib}, $\inspbon$.
Through the (currently largely uncertain) \LIMRI{} merger rate, these two quantities are directly proportional to the total number of detections and the fraction of detections from systems that do not merge during the mission lifetime, respectively.

The paper is organised as follows. In Sec.~\ref{sec:ndet}, we introduce the \textit{sensitivity time-volume} and the \textit{\ib}{}, and show their connection to the expected number of detections through the intrinsic merger rate of the binaries. In Sec.~\ref{sec:wav}, we outline the waveform model used in this work, based on the \FEW framework, and describe the extensions developed for this study. Section~\ref{sec:results} presents our results, with their implications discussed in Sec.~\ref{sec:impl}. Finally, Sec.~\ref{sec:concl} summarises our conclusions and provides pointers to future work.

Throughout this paper we use geometric units, in which $G = c = 1$. 
We adopt Planck18 Cosmology~\citep{planck2018} as implemented in \texttt{astropy}~\citep{astropy2013, astropy2018, astropy2022}.

\section{LISA survey's observational reach}
\label{sec:ndet}

Let us consider the cosmic population of \LIMRI{}s in which a given binary at redshift $z$ -- and corresponding \textit{luminosity} distance $D_L(z)$ -- is  characterised by (source-frame) masses $m_{1,2}$, spin vectors $\vec{\chi}_{1,2}$ and eccentricity $e$ (at some reference time). 
Other mass parameters that we will consider are the total and reduced mass, $M = m_1+m_2$ and $\mu = m_1 m_2/M$, respectively, the mass ratio $q \equiv m_2/m_1 \le 1$ and the symmetric mass ratio, $\eta \equiv \mu/M = q/(1+q)^2$.

We will denote by $\bm\theta = \left\{ m_1, m_2, \vec{\chi}_1,  \vec{\chi}_2, e\right\}$ the vector describing the physical parameters of the binary. 

In this Section we derive the two key metrics that we use to assess the LISA's ability to observe \LIMRI{}s:
\begin{itemize}
    \item The \text{sensitivity time-volume}, $\VT(\bm\theta)$: it provides the hyper-volume (the accessible duration of the source lifetime multiplied by the volume of the universe) for a source with physical parameters $\bm\theta$ that LISA can reach. The total number of \LIMRI{}s characterised by $\bm\theta$ detectable by LISA is just the sensitivity time-volume multiplied by the intrinsic merger rate.
    \item The \ib, $\inspbon$: the ratio of the sensitivity time-volume of systems that merge during the mission over the whole sensitivity time-volume. It provides a measure of the contribution of systems far from plunge to the total number of detections.
\end{itemize}

At a given time $t$, a \LIMRI is at some stage of its evolution, which here we assume to be entirely driven by gravitational radiation reaction, corresponding to a remaining (source-frame) lifetime $\tau$, where
\begin{equation}
    \tau = t_p - t \,,
    \label{eq:tau}
\end{equation}
and $t_p$ identifies a specific ``final" instant of the binary's existence, which we set to be the plunge time.
This choice is driven by the waveform model that we use in this work, see Sec.~\ref{sec:wav}, and corresponds to the latest time at which we can evolve the binary and compute waveforms. 
Note that a different choice of reference time, \textit{e.g.} time of coalescence or time at which the GW amplitude reaches its peak, does not affect the results presented in this paper. 
These times differ by a factor on the order of $\sim 0.1\, [(m_1 + m_2)/10^4\,\Msun]\, \mathrm{sec}$, compared to the time-period (in general many years) over which a binary is observable.

The total number of systems detected by LISA can be formally written as (see \textit{e.g.} Ref.~\cite{Gerosa+2019})
\begin{widetext}
\begin{subequations}
\label{eq:Ndet}
    \begin{align} 
        \Ndet & = \int_\Theta \d\bm\theta \int_{0}^{z_\mathrm{M}} \d z \int_0^{\tau_{\mathrm{M}}^{(\mathrm{obs})}} \d\tau^{\prime}\,\rate(\bm\theta, z)\, \frac{dV_c}{dz}\, \frac{1}{1+z}\, p_\mathrm{det}(\bm\theta, z, \tau^{\prime}) \,,
        \label{eq:Ndet_a}
        \\
        & = \int_\Theta \d\bm\theta\, \rate_0(\bm\theta)\, \langle \lambda(\bm\theta) \rangle\, \VT(\bm\theta)\,. 
        \label{eq:Ndet_b}
    \end{align}
\end{subequations}
\end{widetext}
In the previous equation, $\rate(\bm\theta, z)$ is the binary \textit{merger rate}, $\pdet$ is the probability of the instrument to detect a binary, and ${dV_c}/{dz}$ the differential co-moving volume. In going from Eq.~\eqref{eq:Ndet_a} to~\eqref{eq:Ndet_b} we have assumed that the merger rate factorises as $\rate(\bm\theta, z) = \rate_0(\bm\theta)\,\lambda(z)$, where $\rate_0(\bm\theta)$ is the local ($z\approx 0$) merger rate and $\lambda(z)$ encodes the redshift evolution.
At the end of Sec.~\ref{ssec:VT}, we relax this assumption, and we show that for a generic $\rate(\bm\theta, z)$, Eq.~\eqref{eq:Ndet_b} is replaced by an expression of similar structure, given by Eq.~\eqref{eq:Ndet_gen}.

In Eq.~\eqref{eq:Ndet_b} we have recast the total number of detections as the product of the intrinsic merger rate and the \textit{sensitivity hyper time-volume} covered by the instrument, $\VT(\bm\theta)$, whose expression is given by Eq.~\eqref{eq:VT} that we derive in Sec.~\ref{ssec:VT}.
The performance of the instrument in observing \LIMRI{}s as a function of their physical parameters is entirely encoded in $\VT(\bm\theta)$ (multiplied by the ${\cal O}(1)$ factor $\langle \lambda(\bm\theta) \rangle$, see below), and this paper is focused on exploring $\VT(\bm\theta)$.

The dimensionless function $\langle \lambda(\bm\theta) \rangle$, the time-volume-averaged $\lambda(z)$, Eq.~\eqref{eq:lambda_ave}, accounts for the effect of the redshift evolution of the merger rate.
It is normalised in such a way that if the merger rate is constant over the volume of the universe covered by the instrument's survey, $\rate(\bm\theta, z) = \rate_0(\bm\theta)$, then $\langle \lambda(\bm\theta) \rangle\ = 1$. 
For any other astrophysically plausible redshift evolution, $\langle \lambda(\bm\theta) \rangle\approx$ a few at most, see Sec.~\ref{ssec:VT}.

We emphasise that in Eq.~\eqref{eq:Ndet} the quantities that describe the binary, $\bm\theta$ and $\tau$, are defined in the source frame. 
In this paper, we will use the convention that observer- (or, equivalently) detector-frame values are denoted by a superscript ``obs", whereas source-frame values are left unmarked unless stated otherwise. 
A generic mass parameter $m$ in the source frame is related to its detector-frame value by $m^{(\mathrm{obs})} = (1 + z)\, m$, and frequency and time intervals are related by $f^{(\mathrm{obs})} = f/(1 + z)$ and $dt^{(\mathrm{obs})} = (1 + z) dt$.  

For completeness, in Eq.~\eqref{eq:Ndet} we have explicitly shown the integration ranges. 
$\Theta$ is the hyper-volume on the physical parameters $\bm\theta$ that characterise the sources. 
The integration on redshift extends from the local universe ($z = 0$) to some maximum redshift, $z_\mathrm{M}$,  at which this class of sources are first born in cosmic history. 
As we have shown in Fig.~\ref{fig:horizon_vs_M} and we will discuss in more detail in Sec.~\ref{sec:results}, in this mass range LISA is sensitive to sources up to $z  \sim$ a few; therefore, we do not need to be concerned about the specific value of $z_\mathrm{M}$, as long as $z_\mathrm{M} \gg 1$. 
The integration on $\tau$ should cover the entire binary lifetime. 
However, as one goes progressively earlier in the source's evolutionary stage, the detection probability goes rapidly to zero, the exact value of $\tau_\mathrm{M}^{(\mathrm{obs})}$ is irrelevant as long as $\Tobs \ll \tau_\mathrm{M}^{(\mathrm{obs})} \ll H_0^{-1}$, where $H_0$ is the value of the Hubble parameter at the present time, and $\Tobs$ the observation time, which is of order of years.

The remainder of the Section is devoted to the derivation of Eq.~\eqref{eq:Ndet} and the various quantities that we have introduced.  

\subsection{Merger rate}

We have defined the \textit{merger rate}, $\rate(\bm\theta, z)$, as the number of binaries that merge per unit co-moving volume, unit time (local to the source) and unit source parameter hyper-volume:
\begin{equation}
    \rate(\bm\theta) \equiv \frac{\d N}{\d t\,\d V_c\, \d\bm\theta} = \rate_0(\bm\theta)\,\lambda(z) \,.
    \label{eq:rate_merg}
\end{equation}
As measured by LISA, in the detector-time interval $\d t^{(\mathrm{obs})}$, the number of mergers that take place within the redshift interval $z$ and $z+\d z$ is
\begin{equation}
\begin{aligned}
\frac{\d N}{\d t^{(\mathrm{obs})}\,\d z\,\d\bm\theta}
&= \frac{\d N}{\d t\,\d V_c\,\d\theta}\,
   \frac{\d t}{\d t^{(\mathrm{obs})}}\,
   \frac{\d V_c}{\d z} \\
&= \rate(\bm\theta, z)\,
   \frac{1}{1+z}\,
   \frac{\d V_c}{\d z}\,.
\end{aligned}
\label{eq:dNdtobsdz}
\end{equation}

From an astrophysical point of view, it is reasonable to assume that binary formation is in steady state and that the systems' evolution is short compared to the time-scales over which the merger rate may change. 
In fact, any change in the demographics of binaries would take place on cosmological timescales, and the maximum value of the binary lifetime that we consider here is $\tau_\mathrm{M}\sim 10^3\,\mathrm{yr}$. 
It is worth pointing out that these assumptions may be violated if a considerable number of (detectable) sources were formed at $\tau \lesssim \tau_\mathrm{M}$, \textit{e.g.} through black hole captures at small orbital separations de-facto forming binaries already well inside the LISA sensitivity window; however, this additional complication is outside the scope of the current paper and we will not consider it here.

Under the assumption of steady-state formation, the number of binaries with a remaining time-to-plunge between \(\tau\) and \(\tau+\mathrm{d}\tau\) equals the number of binaries that merge within the corresponding source-frame time interval between $t$ and $t + \mathrm{d}t$. Consequently, the merger rate in Eq.~\eqref{eq:dNdtobsdz} can also be interpreted as the number of sources per unit observer-frame time-to-plunge, \(\mathrm{d}\tau^{(\mathrm{obs})}\), rather than per unit observer time, \(\mathrm{d}t^{(\mathrm{obs})}\). This follows because both intervals are related to their respective source-frame intervals by the same time-dilation factor, \((1+z)\), thus, \(\mathrm{d}t/\mathrm{d}t^{(\mathrm{obs})}=\mathrm{d}\tau/\mathrm{d}\tau^{(\mathrm{obs})}=1/(1+z)\). Therefore, Eq.~\eqref{eq:dNdtobsdz} directly connects the intrinsic merger rate to the population of binaries emitting gravitational waves within a given remaining-lifetime interval, as measured by the detector.

\subsection{Signal-to-noise ratio and detection probability}
\label{sec:pdet}

\LIMRI{}s are long-lived sources for LISA, as their time to plunge is
\begin{align}
    \tau^{(\mathrm{obs})} & \approx 4.4\times 10^2\,
    x(e)\,
    \left(\frac{\eta}{10^{-3}}\right)^{-1}\times
    \nonumber \\
    & \qquad\left(\frac{f^{(\mathrm{obs})}}{1\,\mathrm{mHz}}\right)^{-8/3}\,
    \left(\frac{(1+z)\,M}{10^4\,\Msun}\right)^{-5/3}\,
    \mathrm{yr}\,,    
\end{align}
where $x(e)$ is a function of eccentricity~\citep{Peters1964}.

LISA will observe only some fraction of the total number of binaries that radiate in the instrument's sensitivity window during mission operation. 
LISA's sensitivity to a specific system depends on the intrinsic parameters of a binary, $\bm\theta$, its redshift, $z$, \textit{and} $\tau$, or the orbital period (or, equivalently, frequency) at the beginning of the observations, in addition to the sky location, $\hat{\Omega}$, and orientation of the orbital angular momentum (at a reference time), or equivalently, its polarisation and inclination angles, $\psi$ and $\iota$, respectively.

The sky- and polarisation-averaged signal-to-noise ratio (SNR) at which LISA detects a source is~\citep{BabakHewitsonPetiteau2021}:
\begin{equation}
	\langle\rho^2(\bm\theta, z, \tau, \Tobs, \iota)\rangle_{\hat\Omega,\psi} =
	4 \int_{\Delta f^{(\mathrm{obs})}}
	\frac{|\tilde h(f^\prime; \bm\theta, z, \iota)|^2}{S_h(f^\prime)}\d f^\prime\,,
\label{eq:snr_skyave}
\end{equation}
where $\tilde h(f; \bm\theta, z,\iota)$ is the Fourier transform of the signal's strain, see Sec.~\ref{sec:wav}, and $S_h(f)$ is LISA's effective noise spectral density for a 6-link constellation that defines the instrument's sensitivity over the observing window $0.1\,\mathrm{mHz} - 1\,\mathrm{Hz}$~\citep{BabakHewitsonPetiteau2021}. 
$S_{h}(f)$ includes the contributions from the instrumental noise, that we model according to the LISA Science Requirements Document (SciRD)~\cite{SciRD}, see also Refs.~\cite{LISA, BabakHewitsonPetiteau2021, LISA_DSR}, and the \emph{confusion noise}, due to background radiation from unresolved short-period binaries, for which we adopt the analytical fit provided in Sec.~9 of Ref.~\cite{BabakHewitsonPetiteau2021} (different approximations may lead to slightly different levels of confusion noise, see \textit{e.g.} Ref.~\cite{KarnesisEtAl2021}). 

In Eq.~\eqref{eq:snr_skyave}, $\Delta f^{(\mathrm{obs})}$ is shorthand for the relevant frequency range spanned by the signal during the observation.
The integration domain clearly depends on $\bm\theta$ and the source's redshift, $z$, which redshifts radiation, but also on $\tau$ and $\Tobs$, as we have explicitly shown in Eq.~\eqref{eq:snr_skyave}. In fact, $\tau$ and $\Tobs$ determine the portion of the binary's signal that is radiated during the observing period. 
For a specified time-to-plunge $\tau$, the binary follows a particular evolutionary trajectory, and the signal is non-zero only over frequencies swept through by the evolving harmonics of the signal during $\Tobs$ (see Sec.~\ref{sec:wav}). 

\begin{figure}
\centering
\includegraphics[width=0.99\linewidth]{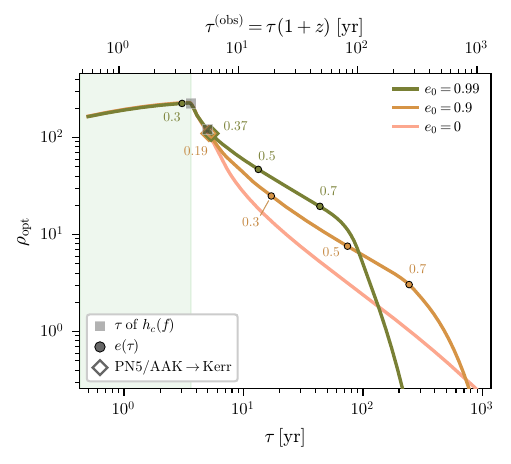}
\caption{The optimal SNR, $\rhoopt$, Eq.~\eqref{eq:snr_opt}, accumulated over an observation of duration $\Tobs = 4\,\mathrm{yr}$ that starts at a time $\tau$ before plunge, for a non-spinning \LIMRI{} with $m_1 = 10^4\,\Msun$ and $m_2 = 10^2\,\Msun$ at $z = 0.11$. The lower (upper) horizontal axis is the source-frame (detector-frame) time to plunge $\tau$ ($\tau^{(\mathrm{obs})} = \tau(1+z)$). Colours denote the eccentricity at $\tau_0 = 10^3\,\mathrm{yr}$: $e_0 = 0$ (pink), $0.9$ (orange) and $0.99$ (green). The pale green band, $\tau^{(\mathrm{obs})} \le \Tobs$, marks the windows that contain the plunge. Filled circles, mark selected values of the instantaneous eccentricity, annotated with that value; grey squares mark the two windows shown in Fig.~\ref{fig:hf_c}; open diamonds mark the handoff from the \AAK{} to the \KerrEq{} model (Sec.~\ref{sec:wav_handoff}), annotated with their handoff eccentricity, $e_\mathrm{hand}$.}
\label{fig:snr_vs_tau}
\end{figure}

To illustrate this point, we have already shown in Fig.~\ref{fig:hf_c} an example of the signal's characteristic strain for a non-spinning \LIMRI{} with $(m_1, m_2) = (10^4, 10^2)\,\Msun$ at $z = 0.11$ for three values of the orbital eccentricity -- we set the eccentricity at a reference time-to-plunge $\tau_0 = 10^3\,\mathrm{yr}$ to be $e_0 \equiv e(\tau = \tau_0) = 0, 0.9, 0.99$ -- and two placements of the observing window of identical duration $\Tobs = 4\,\mathrm{yr}$: one that ends at plunge, and the other one that ends $\approx 1.5\,\mathrm{yr}$ before it.
As a consequence, the SNR accumulated during the observation has a non-trivial dependence on $\tau$ (and eccentricity), which we shown in Fig.~\ref{fig:snr_vs_tau}.
It is evident that for systems close to plunge there is negligible difference between the SNRs obtained for different initial eccentricities, as the sources have had enough time to circularise.
However, already for $\tau^{(\mathrm{obs})} \gtrsim 8\,\mathrm{yr}$ the SNR varies considerably, which, depending on the initial eccentricity, could make a binary detectable tens of years prior to plunge.

In order to quantify the detection probability of a source, we adopt the usual simple SNR-based criterion (see \textit{e.g.} Ref.~\cite{Gerosa+2019}), which is sufficiently accurate for the purposes of this paper.
We consider a binary detectable if the cube root of the mean (over sky location, polarisation and inclination) cube SNR, $\langle \rho^3 \rangle_{\hat\Omega,\psi,\iota}^{1/3}$, exceeds a given SNR threshold, $\rhothr$, \textit{i.e.} 
we set
\begin{equation}
	\pdet = \left\{
	\begin{tabular}{lc}
	$1$ & \qquad $\langle\rho^3 \rangle_{\hat\Omega,\psi,\iota}^{1/3} \ge \rhothr$ \\
	$0$ & \qquad $\langle\rho^3 \rangle_{\hat\Omega,\psi,\iota}^{1/3} < \rhothr $ \\
	\end{tabular}
	\right.\,.
	\label{eq:pdet}
\end{equation}
The fact that \LIMRI{}s are long-lived sources and the instrument moves and changes orientation during the observation time makes the evaluation of $\langle\rho^3 \rangle_{\hat\Omega,\psi,\iota}^{1/3}$ highly computationally intensive. 
Here we use the following approximation to derive $\langle \rho^3 \rangle_{\hat\Omega,\psi,\iota}^{1/3}$ which is sufficiently accurate for the purpose of this paper. 
We compute the sky- and polarisation-averaged SNR for an optimally oriented source (inclination angle $\iota = 0$ or, equivalently, $\pi$) from Eq.~\eqref{eq:snr_skyave}:
\begin{equation}
	\rhoopt(\bm\theta, z, \tau, \Tobs) = \langle \rho^2 (\bm\theta, z, \tau, \Tobs, \iota=0) \rangle_{\hat\Omega,\psi}^{1/2}\,.
	\label{eq:snr_opt}
\end{equation}
The average of $\rho^2$ over $\hat{\Omega}$, $\psi$ and $\iota$ (assuming isotropically located and orientated sources) is related to the average over $\hat{\Omega}$, $\psi$ for an optimally oriented source by $\langle \rho^2 \rangle_{\hat\Omega,\psi,\iota} = (2/5)\,\langle \rho^2 \rangle_{\hat\Omega,\psi,\iota=0}$, see Sec.~7.1.1 in Ref.\cite{BabakHewitsonPetiteau2021}. 
Using the approximation $\langle \rho^3 \rangle_{\hat\Omega,\psi,\iota}^{1/3} \approx \sqrt{3/2}\, \langle \rho^2 \rangle_{\hat\Omega,\psi,\iota}^{1/2}$~\citep{Thorne300Years}, this gives $\langle \rho^3 \rangle_{\hat\Omega,\psi,\iota}^{1/3} \approx 0.77 \rhoopt$, which is what we adopt in Eq.~\eqref{eq:pdet} and to derive the results that we present in Sec.~\ref{sec:results}. 

Equation~\eqref{eq:snr_opt} also allows us to evaluate the (sky- and polarisation-averaged) maximum survey depth of the mission, \textit{i.e.} the highest redshift, $z_\mathrm{max}(\bm\theta)$, or largest luminosity distance, $D_L^{(\mathrm{max})}$, at which a source can be observed.  These quantities also depend on the stage of the evolution of the binaries, so we define
\begin{subequations}
	\label{eq:horiz}
	\begin{align}
		z_\mathrm{max}(\bm\theta) & = \max_{\tau} z_\mathrm{max}(\bm\theta, \tau)
		\label{eq:horiz_z}\,,
		\\
		D_L^{(\mathrm{max})}(\bm\theta) & = \max_{\tau} D_L^{(\mathrm{max})}(\bm\theta, \tau)\,.
		\label{eq:horiz_d}
	\end{align}
\end{subequations}
Throughout the paper we always set $\Tobs = 4\,\mathrm{yr}$, and we adopt a detection threshold $\rhothr = 15$. 
This threshold is higher than the typical threshold used for coalescing binaries in the ground-based literature (typically $\approx 12$) and lower than highly conservative thresholds considered for EMRIs ($\approx 20-30$), see \textit{e.g.} Refs~\cite{ChuaEtAl2017, Speri+2026}. 
This reflects the fact that, although the effective threshold for real-data analyses remains unknown, emerging evidence suggests that this value is achievable even for complex, long-lived signals~\citep{Bandopadhyay+2026}.

\subsection{Sensitivity time-volume}
\label{ssec:VT}

Using Eq.~\eqref{eq:rate_merg}, it is useful to write Eq.~\eqref{eq:Ndet_a} in the form
\begin{align}
    \Ndet & = \int_\Theta \d\bm\theta\, \rate_0(\bm\theta) \int_{0}^{z_\mathrm{M}} \d z \,\lambda(z)\, \frac{dV_c}{dz}\, \frac{1}{1+z}\,\times
    \nonumber \\
        & \qquad \int_0^{\tau_{\mathrm{M}}^{(\mathrm{obs})}} \d\tau^{\prime}\, p_\mathrm{det}(\bm\theta, z, \tau^{\prime}) \,.
        \label{eq:Ndet1}
\end{align}
The integral over $\tau^{(\mathrm{obs})}$ in Eq.~\eqref{eq:Ndet1} plays the role of selecting the portion of the binary evolution, at a given $z$, where the source is detectable. 
We therefore define the (detector-frame) \textit{source visibility-time} as the time interval over which a source is observable, \textit{i.e.} produces an SNR that is above the detection threshold:
\begin{equation}
    \Tvis^{(\mathrm{obs})}(\bm\theta, z) \equiv \int_{0}^{\tau_{\mathrm{M}}^{(\mathrm{obs})}} \d \tau^{\prime}\, p_\mathrm{det}(\bm\theta, z, \tau^{\prime})\,.
    \label{eq:Tvis}
\end{equation}
Notice that $\Tvis^{(\mathrm{obs})}$ depends implicitly on $\Tobs$ through Eqs.~\eqref{eq:pdet} and~\eqref{eq:snr_opt}. 
The upper-limit on the integration time plays an important role. 
Specifically, if we set $\tau_\mathrm{M}^{(\mathrm{obs})} = \Tobs$, we restrict ourselves to systems that merge during the mission duration, \textit{i.e.} $\tau^{(\mathrm{obs})} \le \Tobs$, and therefore $\Tvis^{(\mathrm{obs})}(\bm\theta, z) \le \Tobs$.

Using Eqs.~\eqref{eq:Ndet1} and~\eqref{eq:Tvis}, the total number of detections, Eq.~\eqref{eq:Ndet_a}, can be written as
\begin{equation}
	\Ndet = \int_\Theta \d\bm\theta \rate_0(\bm\theta) \int_{0}^{z_\mathrm{M}} \d z \,\lambda(z)\, \frac{dV_c}{dz}\, \frac{\Tvis^{(\mathrm{obs})}(\bm\theta, z)}{1+z}\,,
	\label{eq:Ndet2}    
\end{equation}
and then expressed in the form given by Eq.~\eqref{eq:Ndet_b}, once one defines the \textit{sensitivity time-volume} as
\begin{subequations}
\label{eq:VT}
	\begin{align}
    		\VT(\bm\theta) & \equiv  \int_{0}^{z_\mathrm{M}} \d z \, 
            \frac{dV_c}{dz}\,  \frac{1}{1 + z}
            \int_0^{\tau_{\mathrm{M}}^{(\mathrm{obs})}} \d\tau^{\prime} \, p_\mathrm{det}(\bm\theta, z, \tau^{\prime})\,,
    		\label{eq:VT_a}
		\\
    		& =  \int_{0}^{z_\mathrm{M}} \d z \,\frac{dV_c}{dz} \frac{\Tvis^{(\mathrm{obs})}(\bm\theta, z)}{1 + z}    \,,
    		\label{eq:VT_b}
		\\
    		& =  \int_{0}^{z_\mathrm{M}} \d z \,\frac{dV_c}{dz} \,\Tvis(\bm\theta, z)    \,,
    		\label{eq:VT_c}
	\end{align}
\end{subequations}
and the time-volume averaged merger rate evolution function as
\begin{subequations}
\label{eq:lambda_ave}
	\begin{align}
    		\langle \lambda(\bm\theta) \rangle & \equiv \frac{1}{\VT(\bm\theta)} \int_{0}^{z_\mathrm{M}} \d z \,\lambda(z)\,\frac{dV_c}{dz}\,  \frac{1}{1 + z} 
            \times \nonumber \\
            & \quad \times \,\int_0^{\tau_{\mathrm{M}}^{(\mathrm{obs})}} \d\tau^{\prime}\,  p_\mathrm{det}(\bm\theta, z, \tau^{\prime}) \,,
		\\
		& = \frac{1}{\VT(\bm\theta)} \int_{0}^{z_\mathrm{M}} \d z \,\lambda(z)\,\frac{dV_c}{dz}\,  \frac{\Tvis^{(\mathrm{obs})}(\bm\theta, z)}{1 + z}\,.
	\end{align}
\end{subequations}
In the case in which the merger rate evolves with redshift, $\langle \lambda(\bm\theta)\rangle$ provides an effective correction to the local merger rate that accounts for both the astrophysical rate's redshift evolution and the redshift interval that contributes to the detection rate through the sensitivity time-volume.
For example, if the redshift evolution tracks the star formation history~\citep{MadauDickinson2014}, $\langle \lambda(\bm\theta)\rangle \approx 1.3\, (4.5)$ at $z=0.1\, (1)$ and reaches a maximum $\langle \lambda(\bm\theta)\rangle \approx 7$ at $z=2.7$.

If the merger rate redshift evolution differs across the \LIMRI's physical parameter space, \textit{i.e.} $\rate(\bm\theta, z)$ can not be factorised as $\rate_0(\bm\theta) \lambda(z)$, one can generalise our analysis and re-write Eq.~\eqref{eq:Ndet_a} as 
\begin{subequations}
    \label{eq:Ndet_gen}
    \begin{align}
	\Ndet & = \int_\Theta \d\bm\theta \int_{0}^{z_\mathrm{M}} \d z \,\rate(\bm\theta, z)\, \frac{dV_c}{dz}\, \frac{1}{1+z}\, \times 
    \nonumber \\
    & \quad \times 
    \int_0^{\tau_{\mathrm{M}}^{(\mathrm{obs})}} \d\tau^{\prime}\,  p_\mathrm{det}(\bm\theta, z, \tau^{\prime})\,
	\label{eq:Ndetgen_a}
	\\ 
		& = \int_\Theta \d\bm\theta \langle \rate(\bm\theta)\rangle\, \VT(\bm\theta)\,,
	\label{eq:Ndetgen_b}
	\end{align}
\end{subequations}
which replaces Eq.~\eqref{eq:Ndet_b}. In Eq.~\eqref{eq:Ndetgen_b}, $\VT(\bm\theta)$ is still given by Eq.~\eqref{eq:VT} and, analogously to Eq.~\eqref{eq:lambda_ave},  the sensitivity time-volume averaged merger rate $\langle \rate(\bm\theta)\rangle$ takes the form
\begin{subequations}
\label{eq:rate_ave}
	\begin{align}
    		\langle \rate(\bm\theta) \rangle & \equiv \frac{1}{\VT(\bm\theta)} \int_{0}^{z_\mathrm{M}} \d z \,\rate(\bm\theta, z)\,\frac{dV_c}{dz}\,  \frac{1}{1 + z} \times
            \nonumber \\
            & \quad \times \,\int_0^{\tau_{\mathrm{M}}^{(\mathrm{obs})}} \d\tau^{\prime\,(\mathrm{obs})}\,  p_\mathrm{det}(\bm\theta, z, \tau^\prime)\,,
		\\
		& = \frac{1}{\VT(\bm\theta)} \int_{0}^{z_\mathrm{M}} \d z\,\rate(\bm\theta, z)\,\frac{dV_c}{dz}\,  \frac{\Tvis^{(\mathrm{obs})}(\bm\theta, z)}{1 + z}\,,
	\end{align}
\end{subequations}
where $\Tvis^{(\mathrm{obs})}(\bm\theta, z)$ is still given by Eq.~\eqref{eq:Tvis}

It is clear from the definition in Eq.~\eqref{eq:VT}, that $\VT(\bm\theta)$ is independent on the intrinsic \LIMRI's astrophysical merger rate and captures the instrument-dependent performance that determines the number of systems observable by LISA. 
Indeed, the advantage of re-writing Eq.~\eqref{eq:Ndet_a} in the form of Eqs.~\eqref{eq:Ndet_b} or~\eqref{eq:Ndetgen_b}, through Eqs.~\eqref{eq:VT} and~\eqref{eq:lambda_ave}, or~\eqref{eq:rate_ave},  is that it allows us to decouple the \LIMRI's unknown intrinsic merger rate from the performance of the instrument which is fully encoded in Eq.~\eqref{eq:VT}.

An additional important consideration for designing and deploying searches, as well as developing the required waveform models, is whether the sensitivity time-volume is dominated by \LIMRI{}s that plunge during the mission or by those that remain observable throughout LISA's operational lifetime and plunge only after the mission ends.
Therefore we introduce the following figure of merit
\begin{equation}
	\inspbon \equiv \frac{\VT(\theta; \tau_{\mathrm{M}}^{(\mathrm{obs})} \gg \Tobs)}{\VT(\theta; \tau_{\mathrm{M}}^{(\mathrm{obs})} = \Tobs)}\,,
    \label{eq:inspbon}
\end{equation}
that we call the \textit{\ib}. 
In the previous equation $\VT(\bm\theta; \tau_{\mathrm{M}}^{(\mathrm{obs})} \gg \Tobs)$ means evaluating Eq.~\eqref{eq:VT} for a choice of the maximum integration time $\tau^{(\mathrm{obs})}$ that is sufficiently large to include all the systems potentially observable by the instrument, whereas $\VT(\bm\theta; \tau_{\mathrm{M}}^{(\mathrm{obs})} = \Tobs)$ restricts the integration limit only to those binaries that plunge within the observation time. 
This is a quantity that satisfies the constraint $\inspbon \ge 1$ and quantifies the advantage, in terms of number of detections, of searching for persistent radiators during the entire mission life-time.

\section{Waveform model}
\label{sec:wav}
We consider a binary system in which a secondary compact object of mass $m_2$ inspirals into a primary IMBH with mass $m_1 \gg m_2$, or equivalently, in which the mass-ratio is
\begin{equation}
    q \equiv \frac{m_2}{m_1} \ll 1\,.
    \label{eq:q}
\end{equation}
Given the component spin vectors $\vec{\chi}_{1,2}$, we define the dimensionless spin parameter of the primary black hole as $a \equiv \pm|\vec{\chi}_1|/m_1^2$, taken positive (negative) for orbits that are prograde (retrograde) with respect to the primary spin.
As we consider binaries with $q \ll 1$ (which complete $\mathcal{O}(q^{-1})$ orbital cycles in-band), we neglect the spin of the secondary object and set $\vec{\chi}_2 = \vec{0}$;
while this parameter contributes $\mathcal{O}(1)\,{\rm rad}$ phase correction over yr-long integration times and is important for parameter estimation~\citep{Skoupy:2025nie,Mathews:2025nyb}, it has negligible impact on SNR calculations~\citep{Burke:2023lno}, which form the basis of the results we present in this work.

The orbital geometry of the binary at a given time is described by the semi-latus rectum $p$, eccentricity $e$ and (cosine of the) inclination of the orbital plane to the binary spin, $x_I$. 
To first order in $q$, these parameters correspond directly to the tangent geodesic in the spacetime of the primary object at that instant.
The position of the secondary object on this geodesic is then fixed by the specification of three Boyer--Lindquist action angles $\Phi_{r, \theta, \phi}$ which encode radial, polar and azimuthal motion respectively~\citep{Schmidt:2002qk} and relate directly to ${r, \theta, \phi}$ coordinates in the primary object's spacetime~\citep{Lynch:2024hco}.

\LIMRI{} systems observable by LISA may retain significant eccentricity in the detection band~\cite{Amaro-Seoane2018, Fragione+2018, ArcaSeddaCapuzzoDolcetta2019, Rasskazov+2020, ArcaSedda+2021, FragioneLoeb2023}.
Motivated by the asymmetric mass-ratios of the binaries that we consider in this work, we therefore model their GW emission with inspiral waveform models in the \textsc{FastEMRIWaveforms} framework (\FEW)~\citep{ChuaEtAl2021, KatzEtal2021, ChapmanBirdEtAl2025, FEWzenodo}, which support high eccentricities $(e \lesssim 0.85)$ even in the strong-field regime and are computationally inexpensive (particularly with hardware acceleration).
As these models do not include the merger-ringdown component of the signal, our results are strictly conservative in this regard, see the discussion in Sec.~\ref{sec:concl}

While \FEW{} houses waveform models suitable for our analyses, the focus of this study is rather different from what \FEW{} was designed for (being tailored to search or parameter-estimation on signals plunging on a timescale of a few years).
Evaluating the sensitivity time-volume, Eq.~\eqref{eq:VT}, requires the SNR, Eq.~\eqref{eq:snr_opt}, of year-long observational windows placed \textit{anywhere} along inspirals in which systems are $\gg \Tobs$ from plunge, repeated across a grid of redshifts for hundreds of choices of system's physical parameters.
This is many orders of magnitude more total waveform evaluations than a standard \FEW{} workflow, and therefore requires us to restructure how the \FEW{} building blocks are deployed. 

In this Section, we describe our waveform generation scheme, detailing which \FEW{} models we employ and the optimisations required to address limitations of \FEW{} relevant to our workflow.
Waveforms generated with our pipeline contain the same physical information as those of \FEW, and every optimisation described below has been validated against the native \FEW{} implementation.
We also caution the reader that all the quantities in this Section (unless explicitly stated) refer to the measured, \textit{i.e.} detector-frame quantities and for ease of notation the super-script ``obs" is omitted throughout.

\subsection{Hybridised trajectory generation}
\label{sec:hybrid-traj}
Here we describe our framework for generating inspiral trajectories given binary parameters, which combines two \FEW{} models as a function of parameter space.
This is necessary because \FEW{}'s relativistic models build upon pre-computed datasets that do not cover the full parameter space, and so cannot support waveforms of duration $\sim 10^3\,{\rm yr}$ in their entirety for the primary mass range $m_1 \in [10^2, 10^5]\,\Msun$ that we consider in this work.
While analytic models exist (one of which we use in this work), they become increasingly inaccurate as the binary separation decreases; by adopting a hybrid prescription, we form an accurate model applicable to the full duration of the inspiral.

Both models we employ in this work operate at adiabatic order, where the inspiral trajectory is obtained by integrating a set of ordinary differential equations:
\begin{subequations}
    \label{eq:odes}
    \begin{align}
    \frac{\d p}{\d t} &= \frac{\eta}{M}\, f_p(p, e, a, x_I)\,,
    \label{eq:traj_odes_p}
    \\
    \frac{\d e}{\d t} & = \frac{\eta}{M}\, f_e(p, e, a, x_I)\,,
    \label{eq:traj_odes_e}
    \\
    \frac{\d \Phi_i}{\d t} & = \frac{1}{M}\, \hat\Omega_i(p, e, a, x_I)\,, \qquad i \in \{r, \phi\}\,,
    \label{eq:phase_odes}
    \end{align}
\end{subequations}
and the forcing functions $f_{p,e}$ can be related to asymptotic energy and angular momentum fluxes at null infinity through flux-balance laws.
The angles $\{\Phi_{r},\Phi_\phi\}$ encode the radial and azimuthal motion of the secondary object respectively, and accumulate according to the (dimensionless) geodesic frequencies $\hat{\Omega}_{r, \phi}$ (which themselves evolve in tandem with $p$ and $e$)~\cite{Schmidt:2002qk, vandeMeent:2019cam}. 
In this work, we restrict ourselves to equatorial orbits (and omit the evolution of $x_I$ and the corresponding polar angle $\Phi_\theta$ from Eqs.~\eqref{eq:traj_odes_p}--\eqref{eq:phase_odes}),  as fast relativistic models for generic (eccentric and inclined) orbits have not yet been implemented in \FEW.
Once these models are available, our framework naturally extends to the generic case.

For the late-inspiral, strong-field portion of the waveform, we use \FEW{}'s fully-relativistic gravitational self-force model for eccentric equatorial orbits~\citep{ChapmanBirdEtAl2025}; in the \FEW{} software notation, this is the \texttt{KerrEccEqFlux} model~\citep{FEWzenodo}.
This model is built upon interpolants of $f_{p,e}$ that are extracted from accurate (but expensive) Teukolsky solutions~\citep{Nasipak_pybhpt_2026}.
It is valid for $|a| \leq 0.999$, $p < 200$ and $e \lesssim 0.9$, although for $p \lesssim 17$, the maximum eccentricity $e_{\rm max}(p)$ gradually tapers as $p$ decreases~\citep{ChapmanBirdEtAl2025}.
To avoid the divergent behaviour of Eqs.~\eqref{eq:traj_odes_p}--\eqref{eq:phase_odes} at the separatrix $p_{\rm sep}$, we terminate trajectory integration at $p_{\rm stop} = p_\mathrm{sep} + \delta$ with $\delta$ chosen to be small.
\FEW{} obtains $p_{\rm sep}(a, e)$ by numerically root-finding the separatrix polynomial of~\citet{Stein:2019buj}; for $a = 0$ this reduces to the simple analytic form
\begin{equation}
    p_\mathrm{sep}(e) = 6 + 2e\,.
    \label{eq:psep}
\end{equation}
Prograde (retrograde) spin moves the separatrix inward (outward) -- for circular orbits, from $p_{\rm sep} = 6$ to $1.45$ ($8.97$) at $a = +0.99$ ($-0.99$) -- while increasing $e$ moves it outward at fixed $a$~\citep{Bardeen+1972}.
Our results are largely insensitive to the choice of $\delta$ (which, for the mass ratios we consider, only impacts the final few cycles of the waveform) and so we opt to retain \FEW{}'s default value of $\delta = 0.002$ for this model~\citep{ChapmanBirdEtAl2025}.
For the same reason, we identify this termination point as the plunge ($\tau = 0$) of the secondary object.

\begin{figure}
    \centering
    \includegraphics[width=0.99\linewidth]{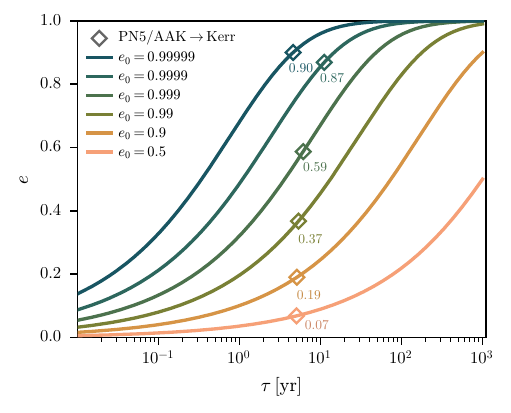}
    \caption{The evolution of the eccentricity as a function of the source-frame time to plunge $\tau$, for the system of Fig.~\ref{fig:snr_vs_tau} (non-spinning BHs of mass $m_1 = 10^4\,\Msun$, $m_2 = 10^2\,\Msun$), for $e_0 = 0.5$ (pink), $0.9$ (golden), $0.99$ (green), $0.999$ (teal), $0.9999$ (blue) and $0.99999$ (dark blue) at $\tau_0 = 10^3\,\mathrm{yr}$. Open diamonds mark the handoff from the \AAK{} to the \KerrEq{} model (Sec.~\ref{sec:wav_handoff}), annotated with the eccentricity $e_\mathrm{hand}$ at which it occurs.}
    \label{fig:e_vs_tau}
\end{figure}

Outside of the domain of the fully-relativistic model, we instead use \FEW{}'s fifth-order post-Newtonian (PN) model, \texttt{PN5}~\citep{Fujita:2020zxe}.
While the PN-expanded $f_{p,e}$ of this model are inaccurate in the strong-field regime, it performs well at the large separations it is applied to in this work~\citep{Isoyama:2021jjd,ChapmanBirdEtAl2025}.
The hybrid model is used only where needed. During integration, the inspiral trajectory is handed over from \texttt{PN5} to \texttt{KerrEccEqFlux} at the first point at which both $p \le 199.9$ (slightly inside the boundary of its domain) and $e \le e_{\rm max}(p)$, yielding $C_0$-continuous $(p, e)$ (and hence $C_1$-continuous phases) at this point. Throughout this work we refer to this as the \textit{handoff}.
More formally, the orbital elements at the handoff are
\begin{equation}
    \begin{aligned}
        p_{\rm hand} & = \max\left\{p : p \le 199.9,\; e \le e_{\rm max}(p)\right\}\,, \\
        e_{\rm hand} & = e(p = p_{\rm hand})\,,
    \end{aligned}
    \label{eq:hand}
\end{equation}
where the maximum runs over the points of the trajectory, \textit{i.e.} it is the first such point reached as $p$ decreases monotonically.
Systems that begin with $p_0 \le 199.9$ and $e_0 \le e_{\rm max}(p_0)$ are evolved with \texttt{KerrEccEqFlux} alone, so have no handoff, though one may consider $p_{\rm hand} = p_0$ and $e_{\rm hand} = e_0$ for completeness.
As the two models agree so precisely at this transition point, we find the precise placement of the $199.9$ threshold has negligible impact on SNR calculations.
As shown in Fig.~\ref{fig:e_vs_tau}, systems circularise efficiently even for highly eccentric initial conditions, so typically $p_{\rm hand} = 199.9$ and the handoff eccentricity $e_{\rm hand} < e_{\rm max} = 0.9$.
For highly eccentric systems the inspiral may reach $p = 199.9$ with $e > e_{\rm max}$.
Rather than discarding these systems, we continue with \texttt{PN5} until $e = e_{\rm max}(p)$, so that $p_{\rm hand} < 199.9$ and $e_{\rm hand} = e_{\rm max}(p_{\rm hand})$, briefly extending the \texttt{PN5} model into the stronger-field regime where it is less accurate.
This scenario (which we refer to as the extended-\AAK{} regime) arises only for $e_0 > 0.9$ at the reference time, and the SNRs of the affected systems should be treated as approximate.
In the results presented in this work, systems in this regime are marked by lilac hatching in Figs.~\ref{fig:vt_m1m2} and~\ref{fig:vtbonus_m1m2}, by a dagger $(\dagger)$ in Table~\ref{tab:e_plunge}, and the trajectory handoff
is represented by open diamonds in Figs.~\ref{fig:snr_vs_tau}, \ref{fig:e_vs_tau} and~\ref{fig:spin}.
\label{sec:wav_handoff}

\begin{table*}
\renewcommand{\thempfootnote}{\fnsymbol{mpfootnote}}
\caption{The eccentricity at plunge, $e_{\rm plunge}$, for selected values of the binary masses and initial eccentricity. Columns 5--9 show $e_{\rm plunge}$ for different values of the initial eccentricity, $e_0$, at time-to-plunge $\tau_0 = 10^3\mathrm{yr}$, for selected values of the individual masses ($m_{1,2}$, first two columns) and associated mass ratio ($q$, third column).
For reference, the fourth column shows the initial value of the semi-latus rectum $p_0 = p(\tau = \tau_0)$ for $e_0 = 0.9$, which is relevant for the choice of trajectory evolution and waveform model, see Sec.~\ref{sec:wav}. \label{tab:e_plunge}}
\begin{ruledtabular}
\begin{tabular}{llllrrrrr}
\colhead{$m_1$} & \colhead{$m_2$} & \colhead{$q$} & \colhead{$p_0(e_0\!=\!0.9)$} & \multicolumn{5}{c}{$e_{\rm plunge}$ for a given $e_0$} \\
\cline{5-9}
{[$\Msun$]} &[$\Msun$] & & & \colhead{0.9} & \colhead{0.99} & \colhead{0.999} & \colhead{0.9999} & \colhead{0.99999} \\
\hline
$10^{2}$ & $0.1$ & $10^{-3}$ & $1057.2$ & 0.0011 & 0.0021 & 0.0032 & 0.0048 & 0.0070 \\
$10^{2}$ & $1$ & $10^{-2}$ & $1867.7$ & 0.0005 & 0.0009 & 0.0015 & 0.0022 & 0.0033 \\
$10^{3}$ & $0.1$ & $10^{-4}$ & $334.5$ & 0.0054 & 0.0095 & 0.0145\footnotemark[2] & 0.0213\footnotemark[2] & 0.0311\footnotemark[2] \\
$10^{3}$ & $1$ & $10^{-3}$ & $594.4$ & 0.0025 & 0.0045 & 0.0069 & 0.0101 & 0.0148\footnotemark[2] \\
$10^{3}$ & $10$ & $10^{-2}$ & $1050.1$ & 0.0012 & 0.0021 & 0.0032 & 0.0048 & 0.0071 \\
$10^{4}$ & $0.1$ & $10^{-5}$ & $106.0$ & 0.0242 & 0.0422\footnotemark[2] & 0.0645\footnotemark[2] & 0.0955\footnotemark[2] & 0.1408\footnotemark[2] \\
$10^{4}$ & $1$ & $10^{-4}$ & $188.1$ & 0.0115 & 0.0200\footnotemark[2] & 0.0305\footnotemark[2] & 0.0450\footnotemark[2] & 0.0657\footnotemark[2] \\
$10^{4}$ & $10$ & $10^{-3}$ & $334.2$ & 0.0054 & 0.0095 & 0.0145\footnotemark[2] & 0.0213\footnotemark[2] & 0.0311\footnotemark[2] \\
$10^{4}$ & $100$ & $10^{-2}$ & $590.5$ & 0.0025 & 0.0045 & 0.0069 & 0.0102 & 0.0149\footnotemark[2] \\
$10^{5}$ & $1$ & $10^{-5}$ & $59.9$ & 0.0510 & 0.0896\footnotemark[2] & 0.1381\footnotemark[2] & 0.2068\footnotemark[2] & 0.3082\footnotemark[2] \\
$10^{5}$ & $10$ & $10^{-4}$ & $106.0$ & 0.0242 & 0.0422\footnotemark[2] & 0.0645\footnotemark[2] & 0.0956\footnotemark[2] & 0.1409\footnotemark[2] \\
$10^{5}$ & $100$ & $10^{-3}$ & $188.0$ & 0.0115 & 0.0200\footnotemark[2] & 0.0305\footnotemark[2] & 0.0450\footnotemark[2] & 0.0658\footnotemark[2] \\
$10^{5}$ & $1000$ & $10^{-2}$ & $332.0$ & 0.0055 & 0.0096 & 0.0146\footnotemark[2] & 0.0215\footnotemark[2] & 0.0314\footnotemark[2]
\end{tabular}
\end{ruledtabular}
\footnotetext[2]{Extended-\AAK{} regime (Sec.~\ref{sec:wav_handoff}).}
\end{table*}

To more efficiently handle the long inspiral trajectories we generate in this work, we modify some aspects of standard trajectory integration with \FEW{} described above.
Of particular importance is that (at adiabatic order) the component masses only enter Eqs.~\eqref{eq:traj_odes_p}--\eqref{eq:traj_odes_e} as scaling factors. 
In terms of the dimensionless time $\tilde t \equiv \eta\, t/M$, they can therefore be re-casted in the form $\d p/\d\tilde t = f_p(p, e)$ and $\d e/\d\tilde t = f_e(p, e)$, yielding a mass-independent solution for the evolution of the orbital elements.
In other words, all binaries with the same initial $(p_0, e_0)$ follow the same parametric curve in the $(p, e)$ plane, at a rate set by $\eta$ and $M$. 
The time to plunge similarly factorises, viz. 
\begin{equation}
   \tau(p_0, e_0; m_1, m_2) = \mathcal{T}(p_0, e_0)\, \frac{M}{\eta}\,,
   \label{eq:lifetime_factor}
\end{equation}
where $\mathcal{T}$ is a dimensionless function of the initial orbital elements alone.
Defining the parameter $\kappa \equiv \eta\, \tau / M$, the initial semi-latus rectum can then be specified directly according to a desired time-to-plunge as $p_0 = \mathcal{T}^{-1}(\kappa,e_0)$.
By constructing an interpolant over $\mathcal{T}^{-1}$, we are then able to efficiently parameterise the binary directly in terms of $\tau$ instead of $p_0$, a more astrophysically intuitive quantity that we use to construct grids of source parameters in our analyses (Sec.~\ref{sec:results}).

In a similar vein, due to the truncation of the equations of motion at adiabatic order, the orbital phases $\Phi_{r, \phi}$ need not be solved in tandem with the orbital element trajectories $\{p(t), e(t)\}$.
Doing so significantly increases the number of steps taken by the ODE solver, which is forced to resolve the evolution of $\Phi_{r, \phi}$ to a sub-radian tolerance over $\gtrsim 10^8$ orbital cycles (recalling that in this work, we evolve the dynamics for upwards of $10^3\,{\rm yr}$).
The increase in the number of steps (which empirically scales as $\sim p_0^{5}$) then forms a substantial bottleneck in our pipeline.
Instead, we construct $\{p(t), e(t)\}$ (the cost of which does not scale appreciably with inspiral duration) and then obtain $\Phi_{r, \phi}$ through quadrature integration of $\hat{\Omega}_{r, \phi}/M$, which yields phase evolutions with sub-milliradian accuracy for all systems we simulate in this work.

\subsection{Waveform generation strategy}
\label{sec:wav_models}
Here we describe our hybrid waveform generation scheme, which builds upon the trajectory generation framework of Sec.~\ref{sec:hybrid-traj} to combine two waveform models within \FEW{} as a function of parameter space.

In the domain of validity of the relativistic trajectory model, we use the corresponding relativistic waveform model, named \texttt{FastKerrEccentricEquatorialFlux} in the \FEW{} software~\citep{ChapmanBirdEtAl2025,FEWzenodo}.
For brevity, we refer to this pairing of trajectory and waveform models as \KerrEq.
As with the trajectory model, it is accurate to adiabatic order --- leading post-adiabatic corrections modify the amplitude at order $\eta \ll 1$, introducing corrections of similar order to the SNR that can be neglected for our purposes.
Despite the limited parameter-space coverage, it is essential for fully-relativistic models to be used in the strong field for accurate SNR estimates: in \cite{ChapmanBirdEtAl2025}, it was shown that semi-relativistic (or `kludge') models can under/overestimate the SNR by more than $40\%$ for equatorial EMRI systems.
The (complex) gravitational waveform is given by the mode-sum
\begin{equation}
    h(t) = \frac{\mu}{D_L} \sum_{\ell mn} H_{\ell m n}(t)\, e^{-i\,\Phi_{mn}(t)}\,,
    \label{eq:h_modes} 
\end{equation}
where the complex mode amplitudes $H_{\ell m n}(t)$ (which depend on the viewing angle of the observer) are functions of $(a, p, e)$ and therefore evolve over the course of the inspiral, and $D_L$ is the luminosity distance to the source.
The $H_{\ell m n}$ are obtained via interpolation over pre-computed grids in $(a, p, e)$ constructed from Teukolsky equation solutions~\citep{Nasipak_pybhpt_2026}, with the same parameter-space coverage as the trajectory flux grids.
The sum runs over the spherical harmonic mode indices $(\ell, m)$ in addition to the radial mode index $n$; as this model (and by extension, our analysis) is restricted to the equatorial plane, the polar mode index $k$ is absent (or equivalently, $H_{\ell mkn} = 0$ for $k \neq 0$).
The carrier phase of each mode is therefore
\begin{equation}
    \Phi_{mn}(t)=m\Phi_\phi(t) + n\Phi_r(t),
\end{equation}
although it should be noted that corrections to these phases are encoded in the argument of $H_{\ell m n}(t)$, which varies on the radiation-reaction timescale $t_{\rm rr} \sim \eta ^ {-1} t_{\rm orb}$, where $t_{\rm orb}$ is the orbital timescale.

In the domain of validity of \KerrEq, we compute the SNR by sampling $h(t)$ at a cadence of $\Delta t = 0.5\,{\rm s}$, Tukey-windowing the resulting time series with shape parameter $\alpha=10^{-2}$ and applying a Fast Fourier transform (FFT).
The choice of $\Delta t = 0.5$ is motivated by the upper cut-off of our analysis band of $1\,{\rm Hz}$; we have verified that this does not introduce significant aliasing over the parameter space we consider in this work.

Outside of the domain of the fully-relativistic models, we instead employ the augmented analytic kludge model (\texttt{AAK})~\citep{BarakCutkler2004, ChuaEtAl2017, KatzEtal2021}, pairing it with the \texttt{PN5} trajectory model of the previous Section (which we refer to in combination as \AAK).
Being fully analytic, the \texttt{AAK} can in principle be evaluated over the full inspiral parameter space and, although its accuracy degrades closer to the separatrix~\citep{KatzEtal2021,ChapmanBirdEtAl2025,Isoyama:2021jjd}, it is sufficiently accurate for the purposes of SNR calculations in the regime in which we apply it here ($p > p_{\rm hand}$, Sec.~\ref{sec:wav_handoff}).
The \texttt{AAK} waveform is constructed from a sum over sideband modes of the quadrupolar emission, with
\begin{equation}
    h(t) = \frac{\mu}{D_L} \sum_{n=1}^{n_{\rm max}} \; \sum_{s = -1}^{1} A^{(s)}_{n}(t)\, e^{-i \left[ n \Phi_r(t) + 2 s\, \tilde\gamma(t)\right]} \, ,
    \label{eq:aak}
\end{equation}
where $\tilde\gamma \equiv \Phi_\phi - \Phi_r$ is the pericentre-precession phase, so that each radial harmonic $n$ consists of three sideband modes ($s = 0, \pm 1$). 
The quadrupolar mode amplitudes $A^{(s)}_n$ are the Peters--Mathews combinations of Bessel functions~\citep{PetersMathews1963}, projected onto the two polarisations through the (constant, since the orbit is equatorial) inclination of the orbital plane relative to the observer.
The number of harmonics retained is $n_{\rm max} = \max\left[4, \lfloor 30\, e_0 \rceil\right]$ by construction~\citep{ChuaEtAl2017}, where $\lfloor\cdot\rceil$ denotes rounding to the nearest integer; we quantify the effect of this truncation in App.~\ref{app:nmax}.

Our analyses require us to compute the SNR over thousands of observation windows per source, and the vast majority of these lie entirely in the regime of large orbital separations where we apply the \AAK.
To optimise our SNR computation for these systems far from plunge, we construct each slowly-evolving sideband mode of Eq.~\eqref{eq:aak} directly in the frequency domain.
Writing $\Psi_{n,s} \equiv n \Phi_r + 2s\tilde\gamma$ for the phase accumulated along harmonic $(n,s)$, we apply the stationary phase approximation~\citep{Hughes:2021exa},
\begin{equation}
    \tilde h_{n,s}(f) \simeq \frac{A_n^{(s)}(t_f)}{\sqrt{|\dot F_{n,s}(t_f)|}}\; e^{\,i\left[2\pi f t_f - \Psi_{n,s}(t_f) - \mathrm{sgn}(\dot F_{n,s})\,\pi/4\right]}\,,
    \label{eq:spa}
\end{equation}
where $t_f$ is the time at which the signal frequency $F_{n,s} = \dot{\Psi}_{n,s}/2\pi$ is equal to $f$.
We then substitute this expression directly into Eq.~\eqref{eq:snr_skyave} to obtain the SNR, avoiding the need to construct each sideband mode in the time domain (which would require evaluation at $ \sim 10^7$ time points, as opposed to a few frequency bins) and reducing computational costs by orders of magnitude~\citep{Speri:2023jte}.
Observational windows that straddle the handoff are assembled in the frequency domain from a stationary phase approximation \AAK{} segment and an FFT \KerrEq{} segment.

In previous work~\citep{Hughes:2021exa,Speri:2023jte}, it has been shown that some higher-order eccentric harmonics exhibit a turnover in their frequency evolution late in the inspiral (due to the behaviour of the azimuthal and radial frequencies as the separatrix is approached) that invalidates the leading-order stationary phase approximation described above.
In this case, Eq.~\eqref{eq:spa} requires higher order corrections, including an Airy function of the second frequency derivative.
For the most eccentric systems, the handoff to the \KerrEq{} model (Sec.~\ref{sec:wav_handoff}) can occur close to this regime, but always precedes the first turnover for the systems we consider in our analysis.
The higher-order correction is never needed for the \AAK{} segments used here and is therefore omitted to minimise computational costs.

One further optimisation allows us to significantly increase the resolution of the grid of redshifts over which we compute SNRs.
Once the source-frame parameters of the binary are specified, changes in redshift impact only the observer-frame component masses and the luminosity distance.
For a given set of source parameters, we first construct the inspiral trajectory and compute waveform mode amplitudes on the sparse nodes returned by the ODE solver, constructing cubic spline interpolants for these functions (as is typically performed in \FEW).
Then, for a given redshift $z$ we rescale the time at each node by $(1+z)$ and adjust the spline coefficients accordingly via the chain rule.
We are then able to construct the waveform as described above, without needing to perform any new trajectory integration or mode amplitude evaluations.
This is an essential component of our methodology, enabling us to rapidly compute SNRs over a range of observation windows and redshifts for inspirals thousands of years in duration.

\section{Results}
\label{sec:results}

We present the results of mapping out the LISA detectable parameter space of \LIMRI{}s focusing on the \textit{sensitivity time-volume}, $\VT(\bm\theta)$,  and the \ib, $\inspbon$, defined in Eqs.~\eqref{eq:VT} and~\eqref{eq:inspbon}, respectively. 

We comprehensively cover the range of physical parameters, $\bm\theta$, characterising \LIMRI{}s. 
Masses span the values $m_1 = 10^2 - 10^5\,\Msun$ and $m_2 = 0.1 - 10^3\,\Msun$, subject to the mass-ratio constraint $10^{-5} \le q \le 10^{-2}$. 
We also explore a wide range of possible residual eccentricity that this class of binaries may retain within the LISA sensitivity band: the ``initial" eccentricity $e_0 = e(\tau = \tau_0)$ at a reference time-to-plunge, Eq.~\eqref{eq:tau}, $\tau_0 = 10^3\,\mathrm{yr}$ is in the range $10^{-5} \le (1 - e_0) \le 1$. 
As radiation reaction efficiently circularises orbits, the maximum value of the eccentricity at plunge, $e_{\rm plunge}$, across the entire $(m_1, m_2, e_0)$ space is $\approx 0.3$, as shown in Table~\ref{tab:e_plunge} (binaries with the same $(p_0, e_0)$ follow the same $(p, e)$ trajectory, Eq.~\eqref{eq:odes}, and therefore reach the same $e_{\rm plunge}$).  
The final parameter that one needs to consider is the spin magnitude of the primary black hole. However, we will show in Sec.~\ref{ssec:spins} that, unlike in the case of EMRIs, the spin's magnitude has a very modest (typically at the $\sim 1\%$ level) impact on the SNR for systems in the mass range considered here. As a consequence, we will set $a=0$ throughout, and do not need to explore the effect of this parameter.

\begin{figure*}[htb!]
    \centering
    \includegraphics[width=0.99\linewidth]{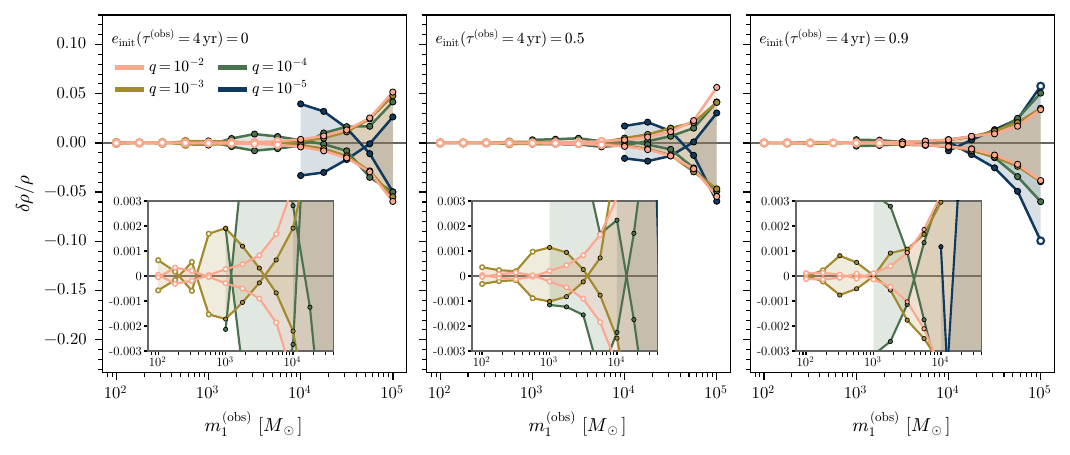} 
    \caption{The effect of the primary BH's spin on the sky- and polarisation-averaged optimal SNR, Eq.~\eqref{eq:snr_opt}. The plot shows $\delta\rho/\rho$, Eq.~\eqref{eq:deltarho_a}, accumulated over the \emph{final} $4\,\mathrm{yr}$ of inspiral, as a function of the detector-frame primary mass $m_1$. For each mass ratio, $q = 10^{-2}, 10^{-3}, 10^{-4}, 10^{-5}$ (pink, golden, green and blue, respectively), the shaded band is bounded by the extremal spins $a = \pm 0.99$; the thin horizontal line marks the non-spinning case, $\delta\rho/\rho = 0$. Filled (open) markers denote binaries computed with the \KerrEq{} model alone (the hybrid model, Sec.~\ref{sec:wav_handoff}). From left to right, the panels correspond to an eccentricity $e_\mathrm{init} = 0, 0.5, 0.9$ at the start of the observation, corresponding to $\tau_\mathrm{init}^{(\mathrm{obs})} = \Tobs = 4\,\mathrm{yr}$. The inset in each panel is a zoom-in of the low-mass region, in which $|\delta\rho/\rho| \lesssim 3\times10^{-3}$.}
    \label{fig:spin_merger}
\end{figure*}

Before we present the results, we need to point out that the low-mass end of the secondary compact objects that we consider here, $m_2 \approx 0.1 - 1\,\Msun$, overlaps with the typical mass of white dwarfs (WDs) and neutron stars, while no BH is currently observationally known in this mass range.
A neutron star in orbit around an IMBH would survive intact throughout the whole insprial and eventually plunge.
Tidal forces affect the final stages of the evolution and leave signatures in the GW phase evolution.
However, because our results depend only on SNR, approximating the evolution of a neutron star as a point particle, therefore covered by the modelling presented in Sec.~\ref{sec:wav}, is sufficient for the scope of this paper and our results equally apply to a secondary object that is a neutron star or a BH.
The situation is radically different if the IMBH's orbital companion is a WD, because depending on the masses of the IMBH and WD (and the WD equation of state), the WD could be tidally disrupted before it plunges into the IMBH~\cite{Rosswog+2009}. 
As a consequence, in Secs.~\ref{ssec:spins}--\ref{ssec:inspbon} we present the results assuming that $m_2$ is always a black hole (or a neutron star), regardless of its mass.
In Sec.~\ref{ssec:wd} we consider the specific case in which the IMBH orbital companion is a WD.

We remind the reader that we set the observing time to $\Tobs = 4\,\mathrm{yr}$ and the detection threshold to $\rhothr = 15$.

\subsection{(Lack of) impact of spins on the sensitivity time-volume}
\label{ssec:spins}

The spin is an important parameter that shapes the signals observed by LISA, and these observations are expected to provide invaluable insights into this fundamental property of black holes, as well as the astrophysical processes that govern the formation and evolution of these systems.
Spins affect the phase evolution of these binaries and need to be taken into account in the design of analysis methods to detect \LIMRI{}s and accurately measure their parameters. 

For extreme-mass ratio inspirals with primary BH masses $\gtrsim 10^5\,\Msun$, spin effects are also known to produce SNR variations of up to a factor of $\approx 10$, with all other binary parameters held fixed.~\citep{ChapmanBirdEtAl2025}. However, this is not the case for binaries in the mass range considered in this paper, as we show below.

\begin{figure*}[htb!]
    \centering
    \includegraphics[width=0.99\linewidth]{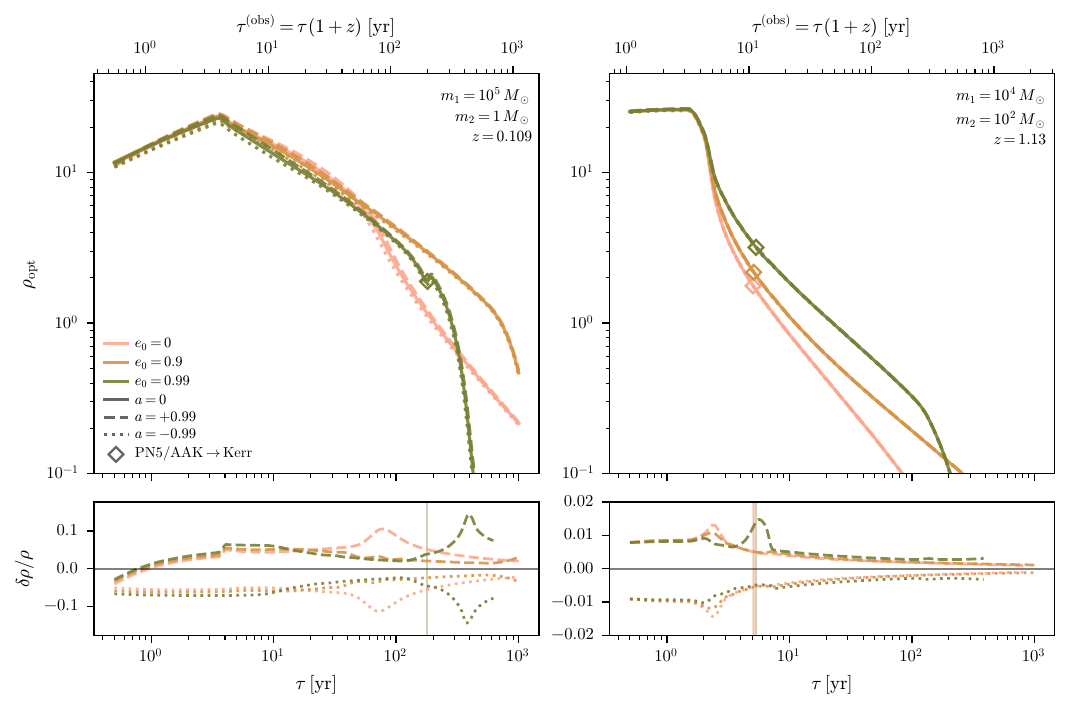} 
    \caption{The impact of the primary BH's spin on the optimal SNR over an observing window of duration $\Tobs = 4\,\mathrm{yr}$, as a function of the time to plunge (source frame, lower horizontal axis; detector frame, $\tau^{(\mathrm{obs})} = \tau(1+z)$, upper horizontal axis), for $(m_1, m_2) = (10^5, 1)\,\Msun$ at $z = 0.11$ (left) and $(m_1, m_2) = (10^4, 10^2)\,\Msun$ at $z = 1.13$ (right). The top panels show $\rhoopt$, Eq.~\eqref{eq:snr_opt}, and the bottom panels $\delta\rho/\rho$, Eq.~\eqref{eq:deltarho_a}; note the different vertical scale of the two bottom panels. Line styles denote $a = 0$ (solid), $a = +0.99$ (dashed) and $a = -0.99$ (dotted), and colours $e_0 = 0$ (pink), $0.9$ (orange) and $0.99$ (green) at $\tau_0 = 10^3\,\mathrm{yr}$. Open diamonds in the top panels, and the corresponding vertical lines in the bottom panels, mark the handoff from the \AAK{} to the \KerrEq{} model (Sec.~\ref{sec:wav_handoff}).
}
    \label{fig:spin}
\end{figure*}

In Fig.~\ref{fig:spin_merger} we show the fractional (sky- and polarisation-averaged optimal) SNR difference
\begin{equation}
    \frac{\delta\rho}{\rho} \equiv \frac{\rhoopt(a)}{\rhoopt(0)} - 1\,,
    \label{eq:deltarho_a}
\end{equation}
where $\rhoopt(a)$ and $\rhoopt(0)$ correspond to the SNR of a binary with BH's spin $a\ne 0$ and $a=0$, respectively, during the final $4\,\mathrm{yr}$ of inspiral as a function of the (detector-frame) primary mass.
We consider parameters in the range $|a|\le 0.99$, $q\in [10^{-5} - 10^{-2}]$ and three different values of the eccentricity at the start of the observations, $e_\mathrm{init} = e (\tau = \Tobs) = 0\,, 0.5\,, 0.9$ (note that $e_\mathrm{init} \ne e_0$).
[We point out that the retrograde \AAK{} waveforms in the current \FEW{} release are affected by a simple bug with a significant impact on the results: the \AAK{} summation assembles the GW harmonics from the \textit{signed} azimuthal phase, which for $x_I = -1$ places the dominant quadrupole at $\sim \forb$, \textit{i.e.}\ half its correct frequency, and with the wrong helicity. We have patched the summation so that, for equatorial retrograde orbits ($x_I = -1$), $\Phi_\phi$ is negated and the strain conjugated in the source frame. With this patch the retrograde \AAK{} model tracks the \KerrEq{} model as the prograde one does, with overlaps $\ge 0.999$ at the handoff.
For completeness, we also highlight that in this Section we always evaluate $\tilde h(f)$ by computing the FFT of $h(t)$ (regardless of the \FEW{} model used) and when we generate waveforms with the \AAK{} model we set $n_{\rm max} = 100$ in Eq.~\eqref{eq:aak}; see App.~\ref{app:nmax} for the effect of harmonic truncation.]

Across the entire parameter space $\left|\delta\rho/\rho\right| \lesssim 6\times 10^{-2}$, and typical values are $\left|\delta\rho/\rho\right| \sim 10^{-2}$ or smaller.
The spread of $\left|\delta\rho/\rho\right|$ changing $a$ from $0.99$ to $-0.99$ varies by more than three orders of magnitude across the mass range, from $\lesssim 10^{-4}$ at $m_1 = 10^2\,\Msun$ (and $\lesssim 2\times 10^{-3}$ for $m_1 \leq 10^3\,\Msun$) to $\approx 5\times 10^{-2}$ at $m_1 = 10^5\,\Msun$, following the fraction of the spin-sensitive strong-field signal that falls inside the LISA band.
Whether a prograde ($a > 0$) or retrograde ($a<0$) binary yields a larger SNR depends on a combination of factors.
At fixed time to plunge, a prograde binary radiates at a slightly higher GW frequency than a retrograde one, by a fractional amount $\propto a\,M f$, and plunges deeper in the strong-field regime.
Where the emitted power is located in the frequency band with respect to the LISA noise curve determines the SNR and is responsible for the ``crossing" of curves that one can see in Fig.~\ref{fig:spin_merger}.

\begin{figure*}[htb!]
    \centering
    \includegraphics[width=0.99\linewidth]{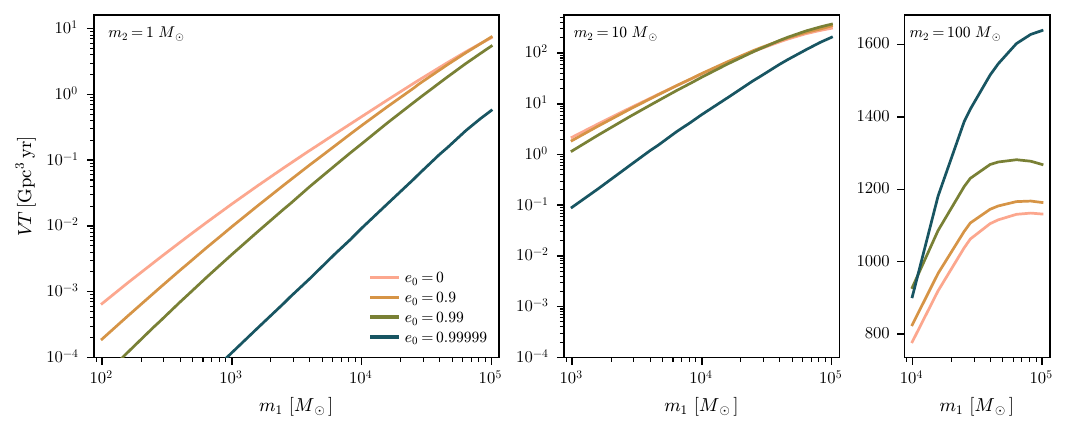} 
    \caption{The sensitivity time-volume, $\VT(\bm\theta)$, as a function of the primary mass $m_1$, for secondary masses $m_2 = 1, 10, 100\,\Msun$ (left, centre, and right panel, respectively) and $e_0 = 0$ (pink), $0.9$ (golden), $0.99$ (green) and $0.99999$ (blue) at $\tau_0 = 10^3\,\mathrm{yr}$. Each panel spans only the range of $m_1$ allowed by $10^{-5}\le q \le 10^{-2}$. Note that the left and centre panels use a logarithmic vertical scale, whereas the right panel a linear one.}
    \label{fig:vt_m1q}
\end{figure*}

These results correspond to systems observed exactly for the final $4\,\mathrm{yr}$ of their lifetime.
One may wonder to what extent these results are affected by the evolutionary stage at which the binary is observed. In order to address this question, we consider two specific systems -- a binary with $(m_1, m_2) = (10^5, 1)\,\Msun$ and one with $(m_1, m_2) = (10^4, 10^2)\,\Msun$ -- that are located at the opposite edge of the mass-ratio range ($q = 10^{-5}$ and $10^{-2}$, respectively), hence characterised by the smallest and largest semi-latus rectum at fixed time to plunge. 
We compute Eq.~\eqref{eq:deltarho_a} as function of the (source-frame) time-to-plunge $\tau$ in the range $\tau \in [0.5, 10^{3}]\,\mathrm{yr}$ as we vary the initial eccentricity, $e_0 = 0, 0.9, 0.99$ at $\tau_0 = 10^{3}\,\mathrm{yr}$, and the primary BH spin $a = 0, \pm 0.99$. The results are shown in Fig.~\ref{fig:spin}. For every value of $\tau$ and $e_0$, $|\delta\rho/\rho| \lesssim 1.5\%$ for the $(10^4, 10^2)\,\Msun$ system and $\lesssim 11\%$ for the $(10^5, 1)\,\Msun$ one, which  supports the conclusion that indeed spins have a modest-to-negligible effect on SNR.
The largest values of $|\delta\rho/\rho|$ are localised in specific points. 
At $\tau \approx 75\,\mathrm{yr}$ (left, $e_0 = 0$) and $\tau \approx 2.4\,\mathrm{yr}$ (right), the signal in the window sits on the knee of the confusion noise (Sec.~\ref{sec:pdet}), where $S_h(f)$ falls steeply with frequency, so the small spin-induced frequency shift produces a disproportionate change in $\rho$; the oscillations of the $e_0 = 0.9$ curves at $\tau \approx 40$--$110\,\mathrm{yr}$ are the same effect acting harmonic by harmonic.
The $e_0 = 0.99$ feature at $\tau \approx 400\,\mathrm{yr}$ (left) is instead numerical: at $n_{\rm max} = 100$ the harmonic sum of Eq.~\eqref{eq:aak} is truncated, and its highest retained harmonic crosses the same knee at this $\tau$; with the sum converged, the residual varies smoothly, from $+3.5\%$ to $+2.2\%$ (prograde) and from $-4.0\%$ to $-2.3\%$ (retrograde).
The narrow prograde feature at the handoff (right) reflects the power-based mode selection of the \KerrEq{} model (Sec.~\ref{sec:wav_models}): a single mode at the threshold is retained only in the prograde run and, because selection is by radiated power rather than SNR, it shifts $\rho$ by $\sim 1\%$.
The step in the prograde residual at $\tau^{(\mathrm{obs})} = \Tobs$ (left) marks the plunge leaving the window: in its final weeks the prograde binary radiates at $20$--$40\,\mathrm{mHz}$, where LISA's noise rises, whereas for retrograde orbits the frequency and amplitude effects cancel.
In general, prograde (retrograde) systems yield a slightly larger (lower) SNR than for $a=0$; however, as we have discussed above, this is not universally true for every choice of $m_{1,2}$ and $\tau$.

In summary, while our results are consistent with the trends that one can extrapolate from Fig.~17 of Ref.~\cite{ChapmanBirdEtAl2025} into the \LIMRI's mass range,
they show that for \LIMRI{} the effect of spins on SNR and as a consequence $\VT$ is typically at most $\sim $ a few $\%$, reaching $\approx 10\%$ in some corners of the $(m_1, m_2)$ plane that are, however, characterised small sensitivity time-volumes, as we will discuss in Sec.~\ref{ssec:vt_res}. Thus, given the modest impact of spin on the SNR and, consequently, on $\VT$, these results motivate our choice to neglect spin, setting $a=0$ unless stated otherwise.
The reader interested in percent level differences, should however be cautioned against ignoring this parameter. 
We note that these conclusions refer to spins aligned with the orbital angular momentum, \textit{i.e.} the equatorial, non-precessing orbits of Sec.~\ref{sec:hybrid-traj}, for which $a$ enters only through the sign convention of Sec.~\ref{sec:wav}. A misaligned primary spin would additionally cause the orbital plane to precess, modulating the signal; we do not model this effect and its impact on the SNR, see however the discussion in Sec.~\ref{sec:concl}.

\subsection{The sensitivity time-volume}
\label{ssec:vt_res}

\begin{figure*}[htb!]
    \centering
    \includegraphics[width=0.99\linewidth]{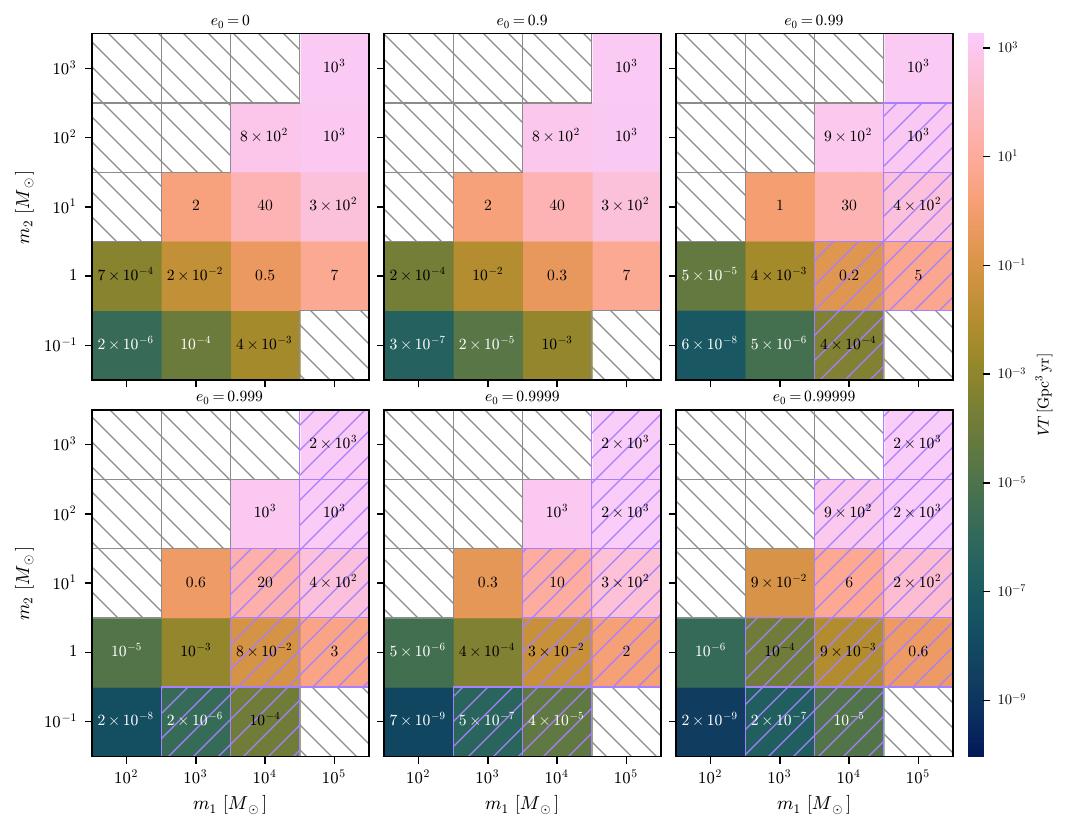} 
    \caption{The sensitivity time-volume,  $\VT(\bm\theta)$ (in $\gpcyr$), across the $(m_1, m_2)$ plane, one panel per initial eccentricity $e_0 = 0, 0.9, 0.99, 0.999, 0.9999, 0.99999$ at $\tau_0 = 10^3\,\mathrm{yr}$ (left to right, top row then bottom row). Cell colour encodes $\VT$ on the shared logarithmic colour bar, and each cell is annotated with its value. Grey-hatched cells lie outside the mass-ratio range $10^{-5} \le q \le 10^{-2}$ and are not computed. Lilac hatching marks cells in the extended-\AAK{} regime, $e > 0.9$ at the handoff (Sec.~\ref{sec:wav_handoff}).
    }
    \label{fig:vt_m1m2}
\end{figure*}

In this Section we explore the behaviour of $\VT(\bm\theta)$, see Eq.~\eqref{eq:VT}.
Our main results are collected in Figs.~\ref{fig:vt_m1q} and~\ref{fig:vt_m1m2}, which show how $\VT(\bm\theta)$ depends on the binary masses $(m_1, m_2)$ and eccentricity. 
Because little is known about the eccentricity distribution of \LIMRI{}s in the LISA band~\cite{Amaro-Seoane2018, Fragione+2018, ArcaSeddaCapuzzoDolcetta2019, ArcaSedda+2021, FragioneLoeb2023}, we explore a wide range of values, from quasi-circular to near-radial orbits. We quote results as a function of the eccentricity $e_0$ at time-to-plunge $\tau_0 = 10^3\,\mathrm{yr}$.
At the end of the Section we provide empirical fits for $\VT(\bm\theta)$ as a function of the binary's parameters.

Figure~\ref{fig:vt_m1q} shows $\VT(\bm\theta)$ as a function of $m_1$ for three values of the secondary mass ($m_2 = 1, 10, 100\,\Msun$) across a range of initial eccentricities. 
The dominant trend is an increase of $\VT(\bm\theta)$ by many orders of magnitude with $m_1$, from $\sim 10^{-6}-10^{-3}\,\gpcyr$ for $m_1 \sim 10^2\,\Msun$ to $\sim 1-10^3\,\gpcyr$ for $m_1 \sim 10^4 - 10^5\,\Msun$, with a weaker dependence on $m_2$ and $e_0$. 
Two effects compound: GWs from more massive primary BHs are emitted at lower frequencies, falling within the more sensitive part of the LISA band and yielding $z_\mathrm{max} \approx 1$--$5$ for the most massive systems with $q \geq 10^{-4}$ (see Fig.~\ref{fig:horizon_vs_M}). Conversely, GWs from less massive primary BHs are emitted at higher frequencies, where LISA's sensitivity degrades; for $m_1 \sim 10^2\,\Msun$, radiation near plunge falls outside the LISA sensitivity band altogether. This limits $D_L^{(\mathrm{max})}$ to $\sim 1$--$500\,\mathrm{Mpc}$ for $m_1 \sim 10^2$--$10^3\,\Msun$. Superimposed on these trends is a monotonic increase in $\VT(\bm\theta)$ with $m_2$ at fixed $m_1$.

The dependence on eccentricity is more subtle, and produces different trends depending on the binary masses.  
For light secondaries ($m_2 \lesssim 10\,\Msun$) circular binaries yield the largest $\VT(\bm\theta)$ at fixed $m_1$;  $\VT(\bm\theta)$ for eccentric systems is suppressed by up to three orders of magnitude as $m_1$ decreases. 
This is due to the fact that for eccentric binaries the radiated power is distributed over a set of harmonics at higher frequencies, in a region of the frequency spectrum where LISA is less sensitive, lowering the SNR and hence $\VT(\bm\theta)$. 
For the heaviest secondaries ($m_2 = 100\,\Msun$, i.e.\ $m_1 = 10^4-10^5\,\Msun$ under our mass-ratio cut) the ordering inverts: circular binaries give the \emph{smallest} $\VT(\bm\theta)$, while moderate eccentricity enhances it by up to a factor of a few. 
Here the power of a circular orbit is concentrated at the low-frequency end of the sensitive band, and higher eccentricity shifts it in a frequency region where the noise is lower. 
The size of the gain depends on where the ``comb" of harmonics falls relative to the noise curve, making it a non-monotonic function of $e_0$.

Figure~\ref{fig:vt_m1m2} maps the same quantity across the $(m_1, m_2)$ plane (within the mass range $10^{-5} \le q \le 10^{-2}$) for six values of the initial eccentricity, from $e_0=0$ to $1-e_0 = 10^{-5}$. 
$\VT(\bm\theta)$ is a steeply increasing function of both masses: within a single panel the extreme corners differ by some nine orders of magnitude.
Reading down in $e_0$, eccentricity erodes $\VT$ throughout the light half of the band and mildly boosts it in the massive corner (the $m_1 = 10^5\,\Msun$ row at the heaviest secondaries), the two-dimensional imprint of the flip discussed above; depending on the mass combination, the change between $e_0 = 0$ and $1-e_0 = 10^{-5}$ ranges from a suppression of up to three orders of magnitude to a mild enhancement.

An important consideration in interpreting the results above is whether sufficiently accurate waveforms are available to enable the detection of sources and the extraction of unbiased physical information from sources across regions of high sensitivity time-volume. To make this point visually clear, in Fig.~\ref{fig:vt_m1m2}, we have hatched in lilac the cells that rely on the extended-\AAK{} regime (Sec.~\ref{sec:wav_handoff}). It is immediately obvious that a significant portion of cells are affected.
The number grows (consistently with expectations) with $e_0$, and for the largest values of $e_0$ considered here essentially all cells are affected.
Truncation of the \AAK{} harmonic sum (App.~\ref{app:nmax}) makes the eccentric values of $\VT$ slight lower bounds: the correction is negligible below $e_0 \simeq 0.99$, reaches a median of $7\%$ and a maximum of $19\%$ at $1-e_0 = 10^{-5}$, and lengthens the visibility time without extending the horizon.

We conclude this Section by providing (empirical) fitting formulae for $D_\mathrm{L}^{\mathrm{(max)}}(\bm\theta)$ and $\VT(\bm\theta)$ that can be useful for future forecast studies that explore the LISA detection rates for \LIMRI{}s.

The horizon distance is well approximated by the function:
\begin{equation}
  D_L^{(\mathrm{max})}(\bm\theta) = 1\, \mathrm{Gpc}\times
  d_1(e_0)\,
  q_3\,M_4^{5/3}\,\mathcal{F}_\mathrm{D}(q, m_1, e_0) \,,
  \label{eq:dmax_fit}
\end{equation}
where we normalise the mass ratio and the primary black mass according to
\begin{subequations}
	\begin{align}
		q_{3} & \equiv \frac{q}{10^{-3}}\,,
		\\
		M_4 & \equiv \frac{m_1}{10^{4}\,\Msun}\,,
    \end{align}
\end{subequations}
In Eq.~\eqref{eq:dmax_fit}, $d_1(e_0)$ is an overall normalisation distance factor that depends (weakly) on eccentricity -- it varies by a factor $\approx 1.7$ across the parameter space -- and measures luminosity distance in units of $\mathrm{Gpc}$, 
\begin{equation}
    d_{1}(e_0) \equiv \frac{D_{L}(e_0)}{1\,\mathrm{Gpc}}\,,
\end{equation}
and the function 
\begin{equation}
  \mathcal{F}_\mathrm{D}(q, m_1, e_0) = 
  \left[\,1+ 0.025 \left(\frac{40}{Y_h(e_0)}\right)
  q_3\,M_4^{5/3}\right]^{-\xi}
  \label{eq:F_D_fit}
\end{equation}
captures the subtle dependency on the eccentricity, which varies depending on the masses of the binary. The slope of the function $\mathcal{F}_\mathrm{D}$ is common to all eccentricities, $\xi = 1.49$, while the constant $Y_h(e_0)$ varies by a factor $\approx 2.3$. The values of the fitting parameters in Eq.~\eqref{eq:dmax_fit} are given in Table~\ref{tab:fits} in App.~\ref{app:fits} where we provide additional details on the applicability of the fitting formulae. The key point to (re-)emphasise here is that, while $D_L^{(\mathrm{max})}$ is a strong function of the masses of the binaries, it depends weakly on the eccentricity of the system (and even less on the spins).

Analogously, we can fit the sensitivity time-volume and we find:
\begin{align}
   \VT(\bm\theta) & = 100\, \gpcyr \times
   \VT_{100}(e_0)\,
    \nonumber \\
    & \quad \times q_3^{s_1(e_0)}
    M_4^{5s_1(e_0)/3}\,
    \mathcal{F}_\mathrm{VT}(q, m_1, e_0)\,,
    \label{eq:VT_fit}    
\end{align}
where 
\begin{equation}
    \VT_{100}(e_0) \equiv \frac{\VT(e_0)}{10^2\,\gpcyr}
\end{equation}
is an overall normalisation, which is a function of the eccentricity. The function $\mathcal{F}_\mathrm{VT}(q, m_1, e_0)$ describes the rather non-trivial effect of the eccentricity,
\begin{widetext}
\begin{equation}
  \mathcal{F}_\mathrm{VT}(q, m_1, e_0) = 
  \left\{\,1+ \left[0.2\left(\frac{5}{Y(e_0)}\right)
  q_3\,
  M_4^{5/3}\right]^{(s_1(e_0)-s_2)/v(e_0)}\,\right\}^{-v(e_0)}\,,
  \label{eq:F_VT_fit}
\end{equation}
\end{widetext}
where $s_2 = -0.318$. The values of the fitting parameters that enter Eq.~\eqref{eq:VT_fit}, $\VT_{100}(e_0)$, $s_1(e_0)$, $Y(e_0)$ and $v(e_0)$ are given in Table~\ref{tab:fits} in App.~\ref{app:fits}, where we provide additional details on the range of validity and use of these fits.

\begin{figure*}[htb!]
    \centering
    \includegraphics[width=0.99\linewidth]{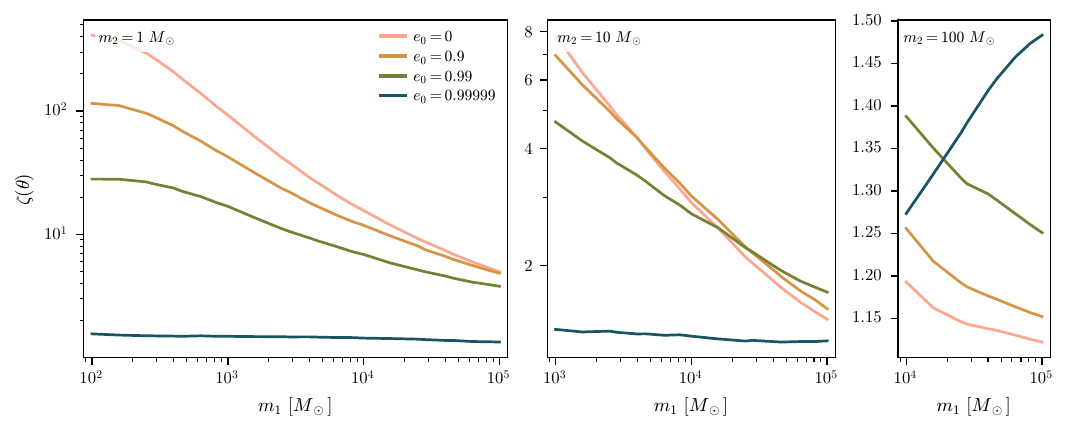} 
    \caption{The \ib{} $\inspbon$, Eq.~\eqref{eq:inspbon}, as a function of the primary mass $m_1$, for secondary masses $m_2 = 1, 10, 100\,\Msun$ (left, centre, and right panel, respectively) and $e_0 = 0$ (pink), $0.9$ (golden), $0.99$ (green) and $0.99999$ (blue) at $\tau_0 = 10^3\,\mathrm{yr}$. As in  Fig.~\ref{fig:vt_m1q} the left and centre (right) panels use a logarithmic (linear) vertical scale.
    }
    \label{fig:vtbonus_m1q}
\end{figure*}

In contrast to the case of $D_\mathrm{L}^{\mathrm{(max)}}(\bm\theta)$, $\VT(\bm\theta)$ can change by up to three orders of magnitude depending on $e_0$. This is due to the fact that the sensitivity time-volume is the integral of the visibility time (that varies greatly depending on the distance and the eccentricity) multiplied by each volumetric distance/redshift shell integrated over the redshift, Eq.~\eqref{eq:VT_c}. 

\subsection{Sensitivity prior to plunge}
\label{ssec:inspbon}

The sensitivity time-volume is the result of integrating over the entire observable lifetime of systems within each detectable redshift shell and does not, by itself, reveal which stage of the inspiral dominates it. 
For search design this distinction is essential: it matters whether $\VT(\bm\theta)$ is dominated by \LIMRI{}s merging within the mission lifetime ($\tau^{(\mathrm{obs})} \le \Tobs$) or whether inspirals still far from coalescence contribute significantly.

\begin{figure*}[htb!]
    \centering
    \includegraphics[width=0.99\linewidth]{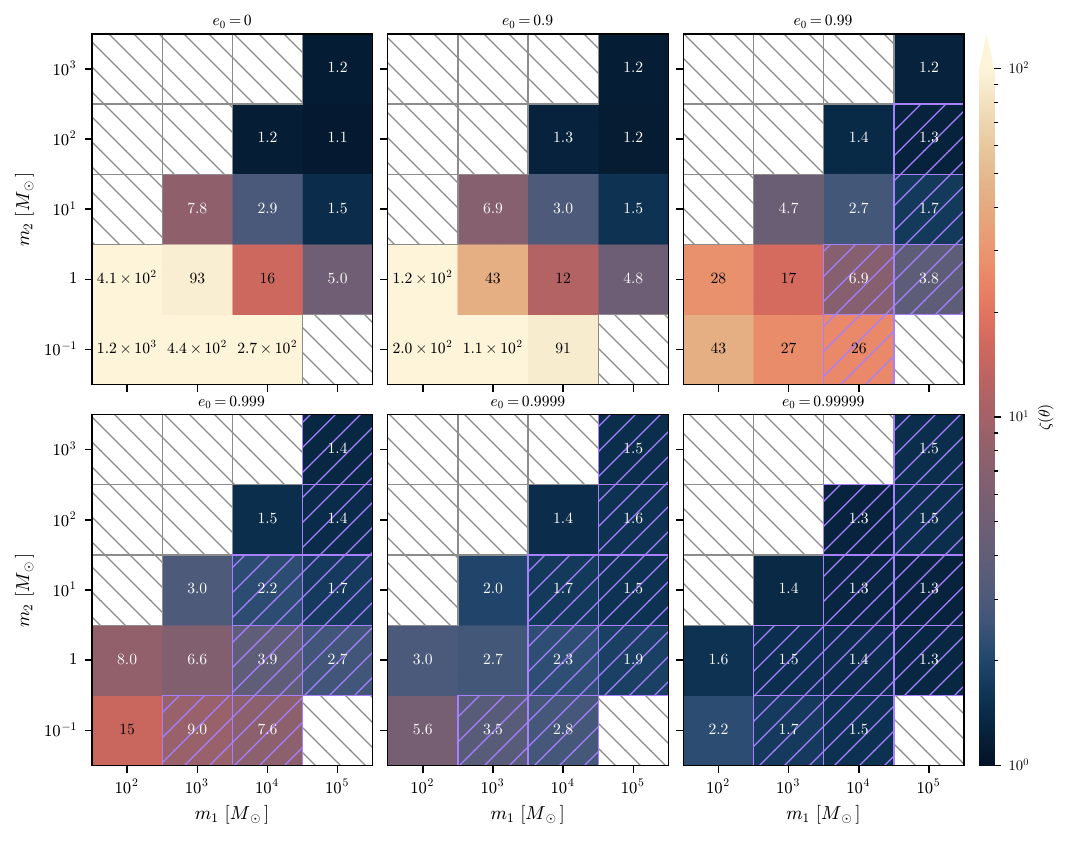} 
    \caption{The \ib{} $\inspbon$, Eq.~\eqref{eq:inspbon}, across the $(m_1, m_2)$ plane, with the same panels, cells and hatching as Fig.~\ref{fig:vt_m1m2}. The logarithmic colour scale runs from $\inspbon = 1$ (dark blue) and saturates at $\inspbon = 100$; annotated values in the saturated cells extend to $\inspbon = 1.2\times10^3$.}
    \label{fig:vtbonus_m1m2}
\end{figure*}

%
%
\begin{figure}
    \centering
    \includegraphics[width=0.99\linewidth]{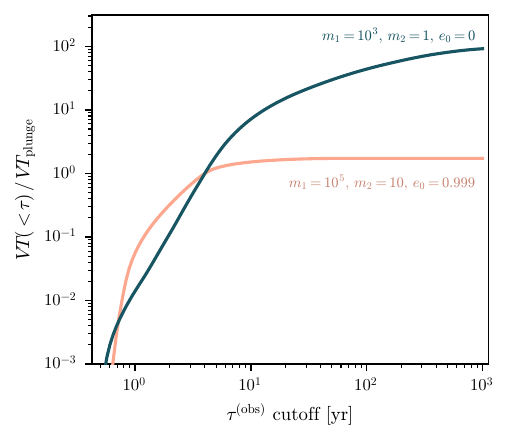}
    \caption{The build-up of the \ib{} $\inspbon$, Eq.~\eqref{eq:inspbon}, as the observational window cover progressively earlier stages of the inspiral, for two contrasting systems: a light, circular binary ($m_1 = 10^3\,\Msun$, $m_2 = 1\,\Msun$, $e_0 = 0$; blue) and a massive, eccentric one ($m_1 = 10^5\,\Msun$, $m_2 = 10\,\Msun$, $e_0 = 0.999$; pink), each labelled in place. We show $\VT(<\tau)/\VT_\mathrm{plunge}$, the ratio between the time-volume accumulated out to a detector-frame time-to-plunge cutoff $\tau^{(\mathrm{obs})} = \tau(1+z)$ and the time-volume from merger-capturing windows alone, as a function of that cutoff. By construction the curves cross unity at $\tau^{(\mathrm{obs})} = \Tobs = 4\,\mathrm{yr}$, and their endpoints are the corresponding values of the \ib, $\inspbon = 91.7$ and $\inspbon = 1.72$, respectively.}
    \label{fig:vtbonus_example}
\end{figure}
%
%

In order to quantify this behaviour, we have defined in Sec.~\ref{ssec:VT} a quantity that we call the \ib, $\inspbon$, the ratio between the total $\VT(\bm\theta)$ and the $\VT(\bm\theta)$ of systems plunging during the mission lifetime.  
By definition $\inspbon \ge 1$, with $\inspbon = 1$ when only plunging systems contribute to $\VT(\bm\theta)$ and $\inspbon \gg 1$ when $\VT(\bm\theta)$ is dominated by systems that will not plunge during mission operation.

Figures~\ref{fig:vtbonus_m1q} and~\ref{fig:vtbonus_m1m2} present $\inspbon$ in the same layouts as Figs.~\ref{fig:vt_m1q} and~\ref{fig:vt_m1m2} (which refer to $\VT(\bm\theta)$).
The central result is that (i) in the areas of the parameter space where $\VT(\bm\theta) \gg 1\,\gpcyr$, $\inspbon \approx 1.1 - 3$,  \textit{i.e.} a significant contribution to the sensitivity time-volume comes from long-lived systems that are far from plunge, and (ii) the region where $\VT(\bm\theta) \ll 1\,\gpcyr$ is characterised by $\inspbon \gg 1$, meaning that in this portion of the $(m_1, m_2)$ plane the sensitivity time-volume is entirely dominated by binaries that do not plunge during mission operation. 
To first approximation, the trend is  that $\inspbon$ is the photographic negative of $\VT(\bm\theta)$: it is the largest in the area of the parameter space  where $\VT(\bm\theta)$ is the smallest, but neglecting \LIMRI{}s that do not plunge during mission operation would significantly decrease the number of LISA detections.

For near-circular light systems ($m_1 \approx 10^2-10^3\,\Msun$, $m_2 \approx 1\,\Msun$, $e_0 = 0$) we find that $\inspbon \sim 10^2$, which implies that LISA's detection capability for these binaries is overwhelmingly for systems far from plunge; a few of the lightest binaries are de-facto detectable at any meaningful astrophysical distance \emph{only} during the early inspiral.  
At the opposite corner of the $(m_1, m_2)$ plane, the more massive \LIMRI{}s ($m_1 \approx 10^4-10^5\,\Msun$, $m_2 \approx 10^2-10^3\,\Msun$) yield $\inspbon \approx 2$ (decreasing to $\inspbon \approx 1.1$ for the circular $m_1 = 10^5\,\Msun$, $m_2 = 100\,\Msun$ binary): even in this case a non-negligible contribution to $\VT$ comes from long-lived systems. 

Figure~\ref{fig:vtbonus_example} illustrates these trends for two specific systems.
The more massive binary in a highly eccentric orbit ($m_1 = 10^5\,\Msun$, $m_2 = 10\,\Msun$, $e_0 = 0.999$) reaches $71\%$ of the asymptotic value of $\inspbon$ for $\tau^{(\mathrm{obs})} \approx 10\,\mathrm{yr}$ and $\inspbon$ asymptotes by $\approx 81\,\mathrm{yr}$: there is little to gain from searching further back.
On the other hand, the sensitivity time-volume, and as a consequence the \ib{} for the light, circular binary ($m_1 = 10^3\,\Msun$, $m_2 = 1\,\Msun$, $e_0 = 0$) keeps increasing out to $\tau^{(\mathrm{obs})} \sim 10^3\,\mathrm{yr}$, and likely beyond, contributing meaningfully across nearly its entire pre-plunge lifetime; it is systems of this kind that require searches for sources very far from coalescence.  

The \ib{} approaches unity across the mass parameter space for the largest values of $e_0$, 
because the GW emission power is radiated at higher harmonics that leave the band well before plunge, so only the near-coalescence signal is detectable.
The eccentricity dependence otherwise mirrors that of $\VT$: eccentricity lowers $\inspbon$ at low $m_2$, but for $m_2 = 100\,\Msun$ (and at the high-$m_1$ end of $m_2 = 10\,\Msun$) the most eccentric systems have the largest inspiral contribution (Fig.~\ref{fig:vtbonus_m1q}), since the same redistribution of radiated energy towards the noise minimum, which increases $\VT(\bm\theta)$, also widens the range of $\tau$ over which they remain above threshold.

\subsection{A white dwarf as orbital companion}
\label{ssec:wd}

So far, we have always assumed the constituents of a \LIMRI{} to be BHs, regardless of their mass. 
In this Section we consider a different, specific case in which a WD orbits an IMBH.
Tidal forces exerted by an IMBH can disrupt a WD well before it reaches plunge~\citep{Rosswog+2009}, with potentially profound implications for the figures of merit that have been central to our discussion.
This scenario is also of broader interest in the context of electro-magnetic transient phenomena such as TDEs, QPEs and other unusual transients (some of which have been shown to be consistent with the tidal stripping of a WD by an IMBH)~\cite{Miniutti+2019, Giustini+2020, Arcodia+2021, Arcodia+2024, Nicholl+2024, Chakraborty+2025, Hernandez-Garcia+2025,King2020, King2022, King2023, Franchini+2023, LinialMetzger2023, LinialMetzger2024, Zhou+2024, Yao+2025, Zhou+2025a, Zhou+2025b, Yao2Quataert2026,King2020, Rosswog+2009, MacLeod+2014, MacLeod+2016, Eyles-Ferris+2026,Wang+2022}, with the potential of LISA contributing to multi-messenger observations, see \textit{e.g.} Ref.~\cite{Sesana+2008}.

To emphasise the fact that we consider WDs as orbital companions of an IMBH, we will set $m_2 \equiv \mwd$ for the remainder of this Section, and we will explicitly refer to these systems as WD-IMBHs.
We denote by $\tau_0 \equiv 10^3\,\mathrm{yr}$ the point-particle reference life-time introduced in Sec.~\ref{sec:ndet}, although the true remaining (source-frame) lifetime of the WD, if disrupted, is $\le \tau_0$. The initial eccentricity $e_0 \equiv e(\tau=\tau_0)$ retains its point-particle definition throughout this Section: it labels and seeds the trajectory via the BH-like evolution.

A WD orbiting a BH is tidally disrupted at tidal radius,
\begin{equation}
    R_\mathrm{td} = \left(\eta_\mathrm{td}^2\, \frac{m_1}{\mwd}\right)^{1/3} \Rwd\,,
    \label{eq:Rtd}
\end{equation}
where $\Rwd$ is the (unperturbed) WD radius and $\eta_\mathrm{td}$ is an order-unity parameter which encapsulates the dependence of the disruption condition on the WD's internal structure (its equation of state); we set $\eta_\mathrm{td} = 1$ throughout. 
We adopt the WD mass--radius relation~\citep{Nauenberg1972}
\begin{equation}
    \begin{aligned}
        \Rwd
            &= 8\times 10^8\,\mathrm{cm}
               \left(\frac{M_\mathrm{ch}}{\mwd}\right)^{1/3}
               F(\mwd)^{3/4}\,,\\
        F(\mwd)
            &\equiv 1-\left(\frac{\mwd}{M_\mathrm{ch}}\right)^{4/3}\,,
    \end{aligned}
    \label{eq:wd_mr}
\end{equation}
where $M_\mathrm{ch} = 1.44\,\Msun$ is the Chandrasekhar mass. 

A bound geodesic of semi-latus rectum $p$ and eccentricity $e$ has pericentre distance $r_p = pM/(1+e)$. 
In our analysis we assume for simplicity that complete disruption occurs whenever $r_p \le R_\mathrm{td}$, de-facto imposing that at the first pericentre passage for which this condition is satisfied the binary ceases to exist instantaneously. 
This ansatz, combined with Eqs.~\eqref{eq:Rtd}--\eqref{eq:wd_mr}, provides the semi-latus rectum at disruption,
\begin{equation}
    \begin{aligned}
        p_\mathrm{td}(e)
            & = k\, (1+e)\,\eta_\mathrm{td}^{2/3}\, F^{3/4}(\mwd)\,
            \times
            \nonumber \\
            & \quad
                \left(\frac{m_1}{\Msun}\right)^{1/3}\left(\frac{\mwd}{\Msun}\right)^{-2/3}\left(\frac{M}{\Msun}\right)^{-1}
    \end{aligned}
    \label{eq:ptd}
\end{equation}
where $k = 6.118\times 10^3$ is a dimensionless constant.

As $p$ decreases monotonically during the inspiral, we terminate the evolution at
\begin{equation}
    p_\mathrm{stop}(e) = \max\bigl[\, p_\mathrm{td}(e),\, p_\mathrm{sep}(e) + \delta \,\bigr]\,,
    \label{eq:pstop}
\end{equation}
where $p_\mathrm{sep}(e) + \delta$ is the termination point used in the BH case, see Eq.~\eqref{eq:psep} and surrounding discussion.
Within our assumptions, when $p_\mathrm{td} < p_\mathrm{sep}$, a WD-IMBH evolution is indistinguishable from the one describing a point particle merger; in the opposite case the WD is tidally disrupted.

\begin{figure*}[htb!]
    \centering
    \includegraphics[width=0.99\linewidth]{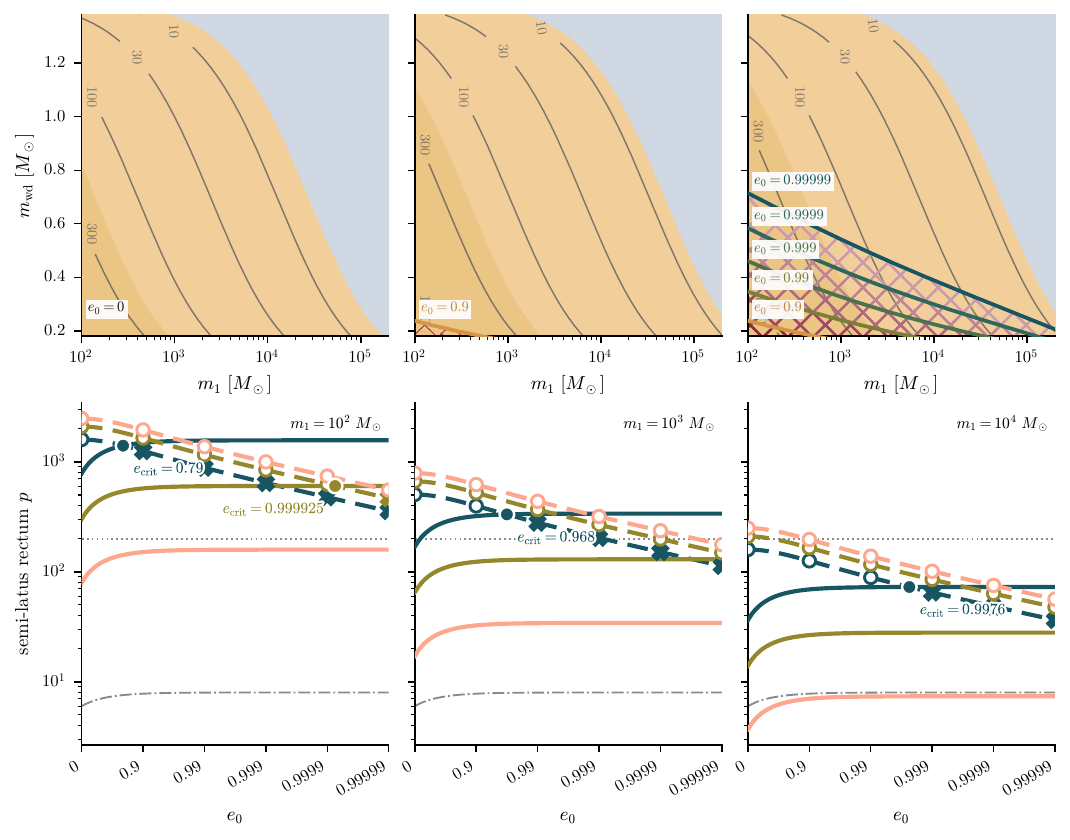}
    \caption{\textit{Top panels:} Partition of the $(m_1,\mwd)$ plane according to the fate of a WD orbiting an IMBH.
    In the pale blue-grey region, $p_\mathrm{td}\le p_\mathrm{sep}$, see Eqs.~\eqref{eq:psep} and~\eqref{eq:ptd}, and a WD inspirals onto an IMBH all the way to plunge. The orange region satisfies the condition $p_\mathrm{sep}<p_\mathrm{td}$ and a WD is tidally disrupted; pale (dark) orange shades correspond to disruption on the \KerrEq{} branch, $p_\mathrm{td}\le199.9$ (on the \AAK{} branch, $p_\mathrm{td}>199.9$, such that the evolution is \AAK{}-only). These regions are visually identical for $e \geq 0.9$.
    Thin grey lines mark contours of constant semi-latus rectum, $p=10,30,100,300,1000$, from which $p_\mathrm{td}$ can be read for any $(m_1,\mwd)$. The left and centre panels show $e_0=0$ and $0.9$, respectively, at $\tau_0=10^3\,\mathrm{yr}$. In the centre panel, the orange curve marks the locus $p_0=p(\tau=\tau_0)=p_\mathrm{td}$; the crimson $\times$-hatched region below it is empty, its systems having been disrupted before $\tau_0$. The right panel overlays this locus for $e_0=0.9$, $0.99$, $0.999$, $0.9999$ and $0.99999$.
    \textit{Bottom panels:} Disruption semi-latus rectum (solid lines), see Eq.~\eqref{eq:ptd}, and initial semi-latus rectum $p_0 = p(e_0)$ at $\tau_0=10^3\,\mathrm{yr}$ (dashed lines) as functions of $e_0$, for $m_1=10^2$, $10^3$ and $10^4\,\Msun$ (panels left to right) and $\mwd=0.2$, $0.6$ and $1.2\,\Msun$ (blue, golden and pink lines, respectively); the corresponding values of $e_\mathrm{crit}$ are listed in Table~\ref{tab:td_ecrit}. Open circles denote values of $e_0$ considered in Fig.~\ref{fig:td_boxes_plate}, filled circles the crossing that defines $e_\mathrm{crit}$, and crosses grid points beyond it, where no orbit exists. Grey dotted and dash-dotted lines show $p=199.9$ (the nominal handoff between the \AAK{} and \KerrEq{} branches) and the ($a=0$) separatrix, Eq.~\eqref{eq:psep}, respectively.}
    \label{fig:td_landscape}
\end{figure*}

For a given initial eccentricity, $e_0$, whether $p_\mathrm{td}$ is always outside the separatrix throughout the evolution of the binary, or it reaches a point where it lies inside the separatrix, induces a partition of the $(m_1, \mwd)$ plane into two regions,  which are shown in the top panels of Fig.~\ref{fig:td_landscape}: (i) \textit{tidal disruption} ($p_\mathrm{td} > p_\mathrm{sep}$), and (ii) \textit{plunge} ($p_\mathrm{td} \le p_\mathrm{sep}$).

Two important caveats are in order. 
In general, WDs will be subject to tidal heating and tidal stripping prior to complete disruption~\citep{2026PhRvD.114b3028Y, 2026arXiv260724951W}. 
These effects are important in determining the exact phase evolution of a binary (and therefore in the actual analysis and interpretation of the data). However, they can be ignored in determining the SNR at which WD-IMBHs can be detected (assuming they are properly accounted for in the analysis), which is the key quantity that determines the sensitivity time-volume. 
In addition, when tidal disruption occurs close to the last stable orbit, the IMBH spin becomes important in determining both whether the WD is disrupted and where in its orbital evolution the disruption occurs.
This effect is neglected here as it affects time-scales much shorter than $\Tobs$ and therefore it has little impact on the SNR. 
Here we are interested in the opposite limit where $\tau$ is comparable or (much) grater than $\Tobs$ as this is what determines $\VT(\bm\theta)$ and related quantities.

The top panels of Fig.~\ref{fig:td_landscape} shows a third region (which we call \textit{empty}) for $\mwd \lesssim 0.7\,\Msun$. 
This is a portion of the $(m_1, \mwd)$ plane, which depends on $e_0$, where no WD-IMBH can exist as it would have been disrupted before reaching $e_0$ at $\tau_0$. 
There are in fact two competing effects that govern how the disruption condition depends on eccentricity at fixed $\tau$. 
The disruption semi-latus rectum, Eq.~\eqref{eq:ptd}, increases with $e$ (by a factor $\approx 2$ as $e\to1$), since disruption is set by the pericentre distance and a more eccentric orbit at fixed $p$ has a smaller pericentre. 
At the same time, the initial semi-latus rectum $p_0$ at fixed $\tau_0$ descreases with $e_0$, and does so faster.
The two curves intersect at a critical eccentricity, $e_\mathrm{crit}$; for $e_0 > e_\mathrm{crit}$, $p_0 \le p_\mathrm{td}(e_0)$ and a binary cannot exist (a putative binary would have been disrupted before reaching this point).
The bottom panels of Fig.~\ref{fig:td_landscape} show this behaviour and values of $e_\mathrm{crit}$ are provided for a selection of WD and IMBH masses in Table~\ref{tab:td_ecrit}. 

\begin{table}
\caption{Critical eccentricity $e_\mathrm{crit}$ above which $p_0(e_0) \le p_\mathrm{td}(e_0)$ at $\tau=\tau_0 = 10^3\,\mathrm{yr}$, \textit{i.e.}\ a WD has already been disrupted earlier in its evolutionary stage ($\tau > \tau_0$). 
Cells marked with ``---'' correspond to mass combinations for which $1-e_\mathrm{crit} \le 10^{-5}$, the largest value of $e_0$ that we consider in this work. 
\label{tab:td_ecrit}}
\begin{ruledtabular}
\begin{tabular}{lccc}
\colhead{$m_1$} & \multicolumn{3}{c}{$\mwd$} \\
\hline
        & $0.2\,\Msun$ & $0.6\,\Msun$ & $1.2\,\Msun$ \\
$10^2\,\Msun$ & 0.790 & 0.999925 & --- \\
$10^3\,\Msun$ & 0.968 & --- & --- \\
$10^4\,\Msun$ & 0.9976 & --- & --- \\
$10^5\,\Msun$ & 0.999935 & --- & ---
\end{tabular}
\end{ruledtabular}
\end{table}

\begin{figure*}[htb!]
    \centering
    \includegraphics[width=0.99\linewidth]{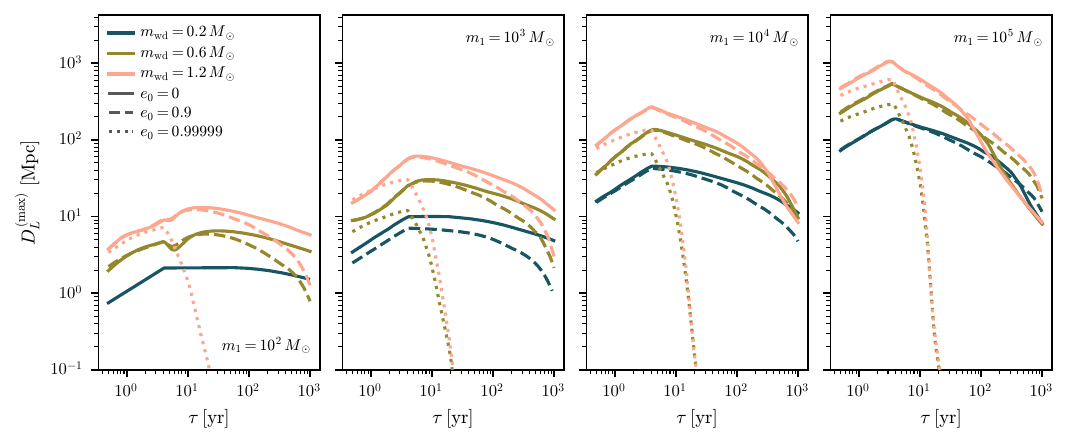}
    \caption{The sky- and polarisation-averaged maximum luminosity distance $D^{(\mathrm{max})}_L$ for WD-IMBHs, as a function of the source-frame time to plunge $\tau$ at which an observation of duration $\Tobs = 4\,\mathrm{yr}$ starts, for $m_1 = 10^2, 10^3, 10^4, 10^5\,\Msun$ (left to right). Colours denote the white dwarf mass, $\mwd = 0.2\,\Msun$ (blue), $0.6\,\Msun$ (golden) and $1.2\,\Msun$ (pink), and line styles the eccentricity at $\tau_0 = 10^3\,\mathrm{yr}$, $e_0 = 0$ (solid), $0.9$ (dashed) and $0.99999$ (dotted). 
    Some of the $(m_1, \mwd, e_0)$ combinations, corresponding to the crimson $\times$-hatched cells in Fig.~\ref{fig:td_boxes_plate}, are not present in the plot as they correspond to eccentricities above $e_\mathrm{crit}$ (see Table~\ref{tab:td_ecrit}).
    Note that in this figure, which refers to WD-IMBHs, the vertical scale is in $\mathrm{Mpc}$, rather than $\mathrm{Gpc}$, which is used in Fig.~\ref{fig:horizon_vs_M}.
    }
    \label{fig:td_horizon}
\end{figure*}

We can now compute all the quantities that we have discussed in the previous Sections, with the simple modification of the trajectory termination condition, Eq.~\eqref{eq:pstop}.
For ease of comparison with the results in previous Sections, we cover the same range of IMBH masses and initial eccentricity, with $\mwd \in \{0.2, 0.6, 1, 1.2\}\,\Msun$.
To complete this grid we include three systems outside the mass-ratio range $10^{-5} \le q \le 10^{-2}$ of Sec.~\ref{sec:results}: $(m_1, \mwd) = (10^2, 1.2)\,\Msun$ ($q = 1.2\times10^{-2}$), and $(10^5, 0.2)$ and $(10^5, 0.6)\,\Msun$ ($q = 2\times10^{-6}$ and $6\times10^{-6}$).

Figure~\ref{fig:td_horizon} shows $D_L^{(\mathrm{max})}(\bm\theta)$ as a function of $\tau$, the time to plunge, for selected values of the WD's mass, $\mwd = 0.2\,, 0.6\,, 1.2\,\Msun$.
At the low-mass end ($m_1 = 10^2-10^3\,\Msun$) of the IMBH mass spectrum, LISA can only reach the nearby Universe, $D_L^{(\mathrm{max})} \sim 2 - 60\,\mathrm{Mpc}$.
It is however worth noticing that, for circular orbits, $D_L^{(\mathrm{max})}$ depends on $\tau$ rather weakly, whereas eccentric systems can only be seen to much smaller distances far from plunge.
The general trend is that (all the other parameters being equal) $D_L^{(\mathrm{max})}(\bm\theta)$ decreases as a function of $e_0$, a consequence of the emitted GW power being radiated at higher frequencies, where LISA's noise is also higher.
For IMBHs of higher mass ($m_1 = 10^4-10^5\,\Msun$), LISA's survey depth reaches from $\approx 40\,\mathrm{Mpc}$ to $\approx 1\,\mathrm{Gpc}$.
At $m_1 = 10^5\,\Msun$, a moderately eccentric binary ($e_0 = 0.9$) is marginally more detectable than a circular one, again a consequence of where the radiated power is located in frequency space with respect to the LISA noise curve.
Here the sensitivity peaks at $\tau \approx \Tobs$, and $D_L^{(\mathrm{max})}$ falls by up to two orders of magnitude as $\tau \rightarrow 10^3\,\mathrm{yr}$.

\begin{figure*}[htb!]
    \centering
    \includegraphics[width=0.99\linewidth]{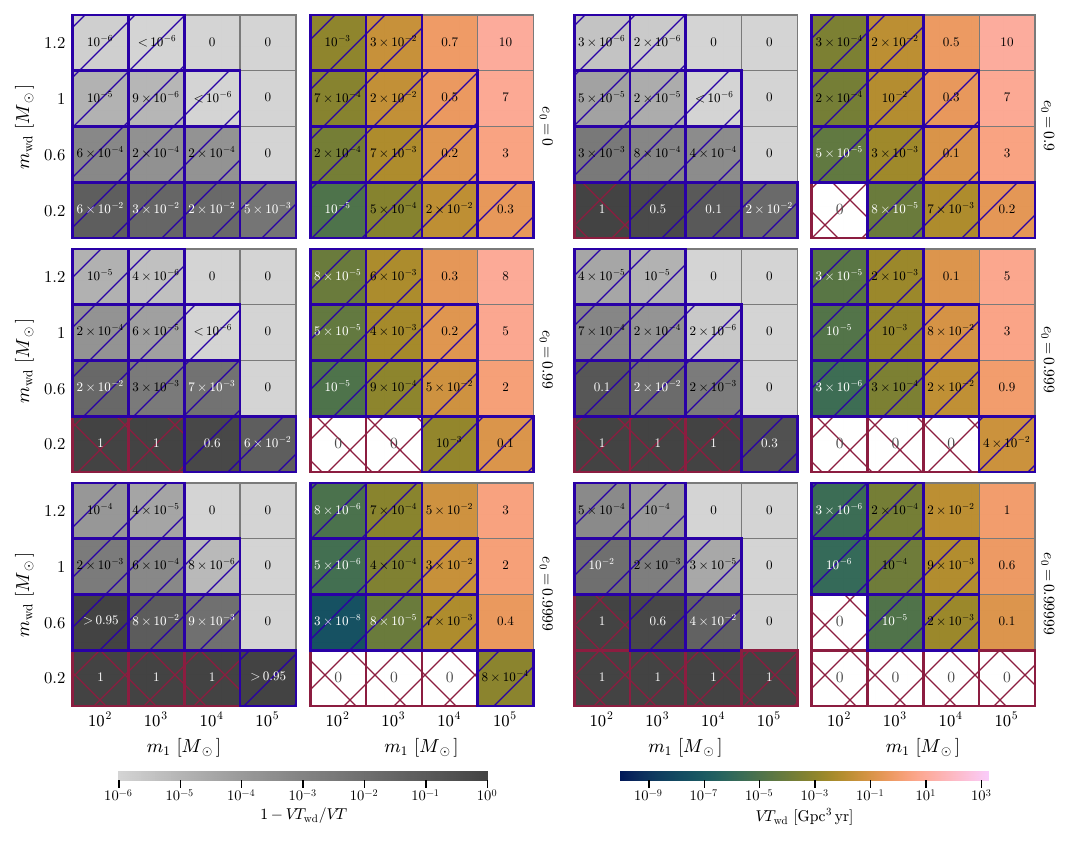}
    \caption{The effect of tidal disruption on the sensitivity time-volume, shown as six pairs of panels, one pair per initial eccentricity $e_0 = 0, 0.9, 0.99, 0.999, 0.9999, 0.99999$ at $\tau_0 = 10^3\,\mathrm{yr}$, labelled on the right-hand edge of each pair and running left to right, top to bottom. Within each pair, the left panel gives the fraction of time-volume lost when the secondary is a white dwarf rather than a black hole, $1-\VT_\mathrm{wd}/\VT$ (grey scale; light grey: no measurable loss, dark grey: total loss), and the right panel the time-volume that survives disruption, $\VT_\mathrm{wd}$ in $\gpcyr$ (colour scale); both colour bars are shared across all pairs and shown at the foot of the figure. Cells span the white dwarf mass $\mwd \in \{0.2, 0.6, 1, 1.2\}\,\Msun$ (vertical axis) and the primary mass $m_1 \in \{10^2, 10^3, 10^4, 10^5\}\,\Msun$ (horizontal axis), and each is annotated with its value. Cell markings record the fate of the system: a plain grey border, the WD survives to plunge; a blue $/$-hatch, the WD is disrupted; a crimson $\times$-hatch, no system exists, $p_0 \le p_\mathrm{td}$ (Table~\ref{tab:td_ecrit}). In the left panels ``$<10^{-6}$'' marks the floor of the grey scale, and ``$1$'' (``$0$'' in the right panels) a cell where a WD-IMBH can not exist.}
    \label{fig:td_boxes_plate}
\end{figure*}

These trends are reflected in the behaviour of the sensitivity time-volume. To emphasise that we consider a WD-IMBH system, we will denote it as $\VT_\mathrm{wd}$, to distinguish it from  $\VT$ discussed in Sec.~\ref{ssec:vt_res}.
The results are summarised in Fig.~\ref{fig:td_boxes_plate}. For $m_1\approx 10^4-10^5\,\Msun$, $\VT_\mathrm{wd}$ reaches $\sim 10^{-1} - 10\,\gpcyr$ when $\mwd \gtrsim 0.6\,\Msun$ for most of the eccentricity range considered here. The sensitivity time-volume decreases very rapidly by several orders of magnitude as $m_1$ decreases (and $e_0$ increases) as most WDs are disrupted and the disruption occurs early on in their evolution. As an example, for $m_1 = 10^3\,\Msun$ and $ \mwd = 0.6\,\Msun$,  $\VT_\mathrm{wd} \sim 10^{-5} - 10^{-2}\,\gpcyr$, depending on the eccentricity.

\section{Implications}
\label{sec:impl}

The results that we have discussed in the previous Section directly translate into estimates of the expected number of LISA detections via Eq.~\eqref{eq:VT} once one assumes an intrinsic merger rate, $\rate(\bm\theta)$.
Merger rate estimates currently span many orders of magnitude,  $\rate \sim 10^{-3} -10^3\,\mathrm{Gpc}^{-3}\,\mathrm{yr}^{-1}$. They reflect the limited understanding of the many astrophysical processes that determine the formation and evolution of \LIMRI{}s and the present lack of observational tracers that constrain many of the model uncertainties surrounding these systems and their environments.

Merger rates in globular clusters at low redshift of \LIMRI{}s consisting of $\approx 10\,\Msun$ BHs in a binary with an IMBH $\sim 10^3\,\Msun$ have been estimated to be in the interval $\rate\sim 1 - 10\,\mathrm{Gpc}^{-3}\,\mathrm{yr}^{-1}$~\citep{Fragione+2018, ArcaSeddaCapuzzoDolcetta2019}, reaching up to $\sim 10^3\,\mathrm{Gpc}^{-3}\,\mathrm{yr}^{-1}$ at $z\approx 2$, which is, however beyond LISA's detection horizon (see Fig.~\ref{fig:horizon_vs_M}). In this mass range $\VT(\bm\theta) \sim 1\,\gpcyr$ for a wide range of eccentricities, decreasing to $\sim 10^{-1}\,\gpcyr$ for only the most extreme eccentricity values, see Figs.~\ref{fig:vt_m1q} and~\ref{fig:vt_m1m2}. Our results therefore imply that LISA could observe $\Ndet\sim 1-10$ systems during its mission lifetime.
Dense star clusters could host \LIMRI{}s with an IMBH $\sim 10^4-10^5\,\Msun$ forming \LIMRI{}s with a BH $\approx 10\,\Msun$ with a corresponding merger rate $\rate \sim 10^{-2} - 10^2 \,\mathrm{Gpc}^{-3}\,\mathrm{yr}^{-1}$~\citep{ArcaSedda+2021}. Given the sensitivity time-volumes in Figs.~\ref{fig:vt_m1q} and~\ref{fig:vt_m1m2} this corresponds to a total number of LISA detections $\Ndet\sim 10^{-1} - 10^4$.

Within the context of the very limited number of unconfirmed (and, in many circumstances highly debated) IMBH candidates currently known, see Fig.~\ref{fig:horizon_vs_M}, these order-of-magnitude estimates show LISA's potential to become a prime observatory for discovery and study of IMBHs, although it is equally clear that detections are by no means guaranteed. 

The results in Sec.~\ref{sec:results} also have implications for the development of the relevant analysis capabilities for these sources.

Firstly, maximising the \LIMRI{} detection yield requires searches that include inspirals far from coalescence, not only systems plunging within the mission lifetime, as we have discussed in Sec.~\ref{ssec:inspbon}. 
Restricting a search to coalescing sources would discard a significant portion (and for the lighter systems the vast majority) of the available time-volume. 
Sensitivity to these earlier evolutionary stages of a binary is also important beyond raw numbers of detections: it offers a view into better preserved imprints (such as eccentricity) of the astrophysical formation processes of these systems.

Secondly, the $\VT(\bm\theta)$ and $\inspbon$ maps, Figs.~\ref{fig:vt_m1m2} and~\ref{fig:vtbonus_m1m2}, identify where further waveform-modelling effort is required. With the exception of near-circular orbits, much of the $(m_1, m_2)$ plane that dominates $\VT(\bm\theta)$ is not yet comprehensively covered by existing accurate \FEW{} implementation (which has required us to devise  tools to combine and interface the \AAK{} to \KerrEq{} models, see Sec.~\ref{sec:wav_handoff}).
Searches for sources in this area of the parameter space, and the characterisation of the physics that a detection would enable, therefore hinge on further extension of faithful eccentric waveform models across these regions. 

\section{Conclusions}
\label{sec:concl}

We have considered GWs from intermediate mass-ratio ($\sim 10^{-5} - 10^{-2}$) inspirals of a low-mass compact object orbiting an intermediate-mass ($\sim 10^2 - 10^5\,\Msun$) BH with the goal of mapping out LISA's sensitivity to this class of sources and therefore guide the development of analysis approaches, waveform models and theoretical astrophysics investigations. 
In fact, \LIMRI{}s have received little attention so far, but given the observational challenge in unveiling IMBHs, LISA may provide important capability to explore the observational ``black hole desert" and shed new light on the astrophysical processes and environments connected to these sources, with the potential for important discoveries.
Given the many astrophysical uncertainties about relevant astrophysical processes, and the fact that LISA may observe systems far from plunge and at a sufficiently early stage of their binary evolution, we consider a wide range of orbital eccentricity for systems radiating in the LISA sensitivity window, spanning from quasi-circular orbits, to highly eccentric binaries (up to an eccentricity $1-e_0\approx 10^{-5}$ at a putative $10^3\,\mathrm{yr}$ from plunge). 
Considering that \LIMRI{}s' mass-ratios are $\ll 1$, we use the most accurate waveforms available in the \textsc{FastEMRIWaveforms}{} framework (\FEW), which we have adapted to the requirements of this study, specifically the need to cover $\sim 10^3\,\mathrm{yr}$ of binary evolution before a source merge.
We have considered two figures of merit for this study: the sensitivity time-volume, which is directly proportional to the total number of expected LISA detections, and the \ib{} which quantifies the fraction of the sensitivity time-volume coming from systems that are long-lived (\textit{i.e.} they do not merge during the observation period) compared to the time-volume of binaries that plunge during the mission lifetime.

Our main findings are as follows:
\begin{itemize}
    \item The sensitivity time-volume varies by many orders of magnitude depending on a \LIMRI's component masses, reaching $\sim 1 - 10^3\,\gpcyr$ ($\sim 10^{-2} - 10^3\,\gpcyr$) for the more massive IMBHs with $m_1 = 10^5\,\Msun$ ($ 10^4\,\Msun$) and sharply decreasing to $\sim 10^{-4} - 1\,\gpcyr$ for lighter primary BHs with $m_1 = 10^3\,\Msun$; the many orders-of-magnitude range in sensitivity time-volume for fixed primary mass is due to both the mass of the orbital companion and the eccentricity of the binary. On the other hand, we find that the sensitivity time-volume is largely insensitive to the IMBH's spin.
    \item A significant portion of the sensitivity time-volume is coming from long-lived sources that do not plunge during the mission and are observed in an earlier phase of their inspiral. The fraction of the time-volume of non-plunging systems varies from $10\%$ to a factor $\sim 10$ for IMBHs with mass $\approx 10^4 - 10^5\,\Msun$ and $\sim 1 - 10^2$ for less massive IMBHs; for the latter, the sensitivity time-volume is entirely dominated by long-lived systems. 
    \item In the case in which the IMBH's orbital companion is a white dwarf, the time-volumes are $\sim 0.1 - 10\,\gpcyr$, $\sim 10^{-3} - 1\,\gpcyr$ and $\sim 10^{-6} - 1\,\gpcyr$ for an IMBH with mass $10^5$, $10^4$ and $10^3\,\Msun$, respectively. Also in this case a significant (and, in many cases, dominant) portion of the time-volume comes from systems that are far from the end of their lifetime (whether the WD plunges into or is disrupted by an IMBH).
    \item For a given \LIMRI's mass, the depth of the LISA survey is fairly insensitive to the orbital eccentricity, but for how long a binary is visible to LISA during its inspiral phase (and, as a consequence, its impact on the sensitivity time-volume) is a strong function of  eccentricity, in a rather non-trivial way.
    \item These results taken together with existing estimates of merger rates of these binaries suggest that LISA may observe $\Ndet \sim 0.1 - 10^3$ \LIMRI{}s, a potentially unique observational sample of IMBHs and their environments.
\end{itemize}

We conclude by cautioning the reading about the assumptions made in deriving the results above, as well as avenues for further work needed to ensure that the full potential of the LISA survey is realised through the development of analysis methods ahead of the mission's launch.

We have adopted waveforms in which the secondary BH's spin is neglected and in which the IMBH's spin is aligned to the orbital angular momentum (therefore we consider equatorial orbits). 
Both choices are driven by the lack of accurate waveforms for the more generic cases.
The former assumption is likely to have negligible quantitative implications on the main results of this work owing to $q\ll 1$. 
The latter deserves to be assessed in more detail -- and astrophysically there is no prior reason to assume the spin is parallel to the binary's angular momentum -- but it is also likely to have $\mathcal{O}(1)$ impact on the figures of merit considered here. 
In fact, they rely on SNR calculations, and spin-induced precession de-facto changes the inclination and polarisation of a binary throughout the observing period. This introduces an effective averaging over (fixed) possible inclination/polarisation that change the SNR by a factor of order unity. 
Needless to say, in the actual processing of the data, the inclusion of precession will be of paramount importance to detect the signals and, importantly, to recover accurate and unbiased measurements of the parameters.

An additional limitation of the waveform models that we consider here is that they are inspiral-only and do not include any contribution from the merger and ringdown produced during the very final stage of binary coalescence. 
As we have shown in Sec.~\ref{sec:results}, \LIMRI{}s are long-lived systems, whose SNR comes from year-long coherent integration.
It is therefore reasonable to assume that neglecting $\mathcal{O}(10)$ signal cycles that follow the inspiral phase will not change the key findings of this paper in any appreciable way. 
In addition, given the masses of IMBHs considered here, the merger/ringdown portion of the signal lies in a region of the frequency spectrum where the LISA noise curve is higher than during the inspiral phase (or for IMBHs with $m_1 \sim 10^3\,\Msun$, outside the observing band) and the impact on SNR of neglecting the merger/ringdown is further reduced. 

For the specific case in which the IMBH orbital companion is a WD, our simplified model considers complete disruption at the first pericentre passage and a simple prescription for the equation of state of the star as a first step in assessing the potential of LISA observations of these sources. 
Future studies should extend the analysis reported here, in particular with the goal of developing analysis approaches to search for these binaries and include the appropriate ingredients for the complex physical phenomena that enter the evolution and GW emission pertinent to these sources.

Our detectability criteria has been simply based on SNR considerations and by fixing a specific threshold motivated by results from ongoing development of analysis strategies that are relevant, but not specifically tailored to searches for \LIMRI{}s, such as EMRIs~\citep{ChuaEtAl2017, Gair+2008, Babak+2009, Wang+2012, Ye+2024, Strub+2026} and stellar-mass binary black holes~\citep{BandopadhyayMoore2024, Fu+2025, Bandopadhyay+2026}. 
As analysis pipelines for this class of signals are developed and a more rigorous assessment of detection probabilities and thresholds (or other relevant quantities in the context of global LISA data processing) become available, they should be considered to provide more robust estimates of the quantities described in Sec.~\ref{sec:results}.  

As a final remark, next generation ground-based interferometers, such as the Einstein Telescope~\citep{Punturo-ET-2010, Abac-ET-2026} and the Cosmic Explorer~\citep{Evans-CE, Reitze-CE} will be optimally placed to observe \LIMRI{}s where the IMBH is a low-mass one $\sim 10^2-10^3\,\Msun$~\citep{Mandel+2008, Gair+2011, Amaro-Seoane2018}. 
Sources in this mass range are therefore multi-band sources, providing further avenues to extract rich science. This particular aspect is beyond the scope of this paper but deserves to be explored in the future.

\begin{acknowledgments}
We thank A.~I.~Ponte Pérez for helpful discussions, along with his input on the visualisation of our results. We also acknowledge the use of Anthropic's Claude Code and Claude models across multiple generations as coding assistance tools for implementing aspects of the analysis and performing numerical consistency checks. The authors reviewed and verified the analysis code and numerical results, excluding plotting and visualisation code. C.E.A.C-B. and A.V. are supported by UKSA Space Agency grant UKRI971. A.V. acknowledges the support of the Royal Society and Wolfson Foundation. 
The computations for this paper were performed using the University of Birmingham's BlueBEAR High Performance Computing facility.

\end{acknowledgments}

\appendix

\section{Harmonic truncation of the \AAK{} model}
\label{app:nmax}

The \AAK{} harmonic sum, Eq.~\eqref{eq:aak}, is truncated at $n_{\max}=\max(4,\lfloor 30\,e_0\rceil)$, keyed on the initial eccentricity, whereas the harmonic power of an orbit of eccentricity $e$ peaks near $n_{\rm peak}=(1-e)^{-3/2}$.
As a cross-check of the validity of our results, we recompute $\VT$ for the $70$ systems of the $(m_1,m_2)$ grid that include an \AAK{} segment, across all six values of $e_0$, with $n_{\max}=\min(12\,n_{\rm peak},3000)$ and $n_{\rm peak}$ evaluated at the eccentricity of each window.
The remaining $8$ systems of the grid, all at $e_0\le0.9$, lie entirely in the \KerrEq{} regime and are unaffected by construction.
This increases $\VT$ by a median of $3.3\%$ and at most $18.6\%$, for $m_1=10^5\,\Msun$, $m_2=10^3\,\Msun$, $1-e_0=10^{-5}$.
The change is entirely due to the increase in visibility time, Eq.~\eqref{eq:Tvis}: neither the maximum distance nor, equivalently, redshift, Eq.~\eqref{eq:horiz}, change by more than $0.5\%$ in any system, so the added number of harmonics lengthens the interval over which a source is detectable without extending how far away it can be seen.
The system carrying the largest $\VT$ correction has an horizon shift of exactly zero.

The median correction grows monotonically with eccentricity and is negligible below $e_0\simeq0.99$: the median (maximum) difference in sensitivity time-volume, $\Delta\VT(\bm\theta)$ is identically zero at $e_0=0$, where a circular orbit radiates in a single harmonic; $1.4\%$ ($2.4\%$) at $e_0=0.9$; $2.2\%$ ($6.7\%$) at $e_0=0.99$; $3.6\%$ ($6.5\%$) at $e_0=0.999$; $5.1\%$ ($7.2\%$) at $e_0=0.9999$; and $7.2\%$ ($18.6\%$) at $e_0=0.99999$.
The correction is not confined to faint systems: for the $21$ systems with $\VT>100\,\gpcyr$ the median is $1.2\%$, and the largest correction in the grid belongs to one of them.

Since adding harmonics can only add power, the values of $\VT$ reported in the main text are lower bounds.
Re-running a sample of truncated windows at $n_{\max}=12\,000$ changes $\Delta\VT$ by $\leq1.5\%$ for the systems with $\VT>100\,\gpcyr$, so the bound is tight where it matters.

\begin{table*}
\caption{Values of the parameters of Eqs.~\eqref{eq:dmax_fit} and~\eqref{eq:F_D_fit}
(left) and Eqs.~\eqref{eq:VT_fit} and~\eqref{eq:F_VT_fit} (right). $s_2$ and $\xi$ are the same for all values of $e_0$ and are given once, in the ``All'' row;
they are held fixed at the values of joint fits across values of $e_0$ ($s_2$ over all $540$ cells, $\xi$ over the $398$ cells with $e_0\le0.9999$), while $D_L(e_0)$, $Y_h(e_0)$, $\VT(e_0)$, $Y(e_0)$ and $v(e_0)$ are fitted separately for each value of $e_0$.
$s_1(e_0)=s_\infty-A_s(1-e_0)^{1/3}$, with $s_\infty=2.876$ and $A_s=0.635$ fitted jointly over all $540$ cells.
Errors (``err'') are median absolute errors, pooled in the ``All'' row, and ``worst'' is the largest single-cell error among cells
with $\VT>1\,\gpcyr$; the excluded cells carry less than $0.05\%$ of the $\VT$ at each value of $e_0$.
\label{tab:fits}
[With $\xi=1.49$ the horizon scales as $(q\,m_1^{5/3})^{-0.49}$ at high mass.
The $e_0=0.99999$ case is poorly described by Eq.~\eqref{eq:dmax_fit} for any $\xi$
(median error $13$--$16\%$); excluding it, the pooled horizon error is $3.3\%$.]}
\begin{ruledtabular}
\begin{tabular}{lrrrrrcrrrrrrr}
\colhead{} & \colhead{} & \multicolumn{4}{c}{Horizon, Eq.~\eqref{eq:dmax_fit}} &
\colhead{} & \multicolumn{7}{c}{Time--volume, Eq.~\eqref{eq:VT_fit}} \\
\cline{3-6}\cline{8-14}
\colhead{$e_0$} & \colhead{$N$} &
\colhead{$D_L(e_0)$} & \colhead{$\xi$} & \colhead{$Y_h(e_0)$} & \colhead{err} & \colhead{} &
\colhead{$\VT(e_0)$} & \colhead{$s_1(e_0)$} & \colhead{$s_2$} & \colhead{$Y(e_0)$} & \colhead{$v(e_0)$} &
\colhead{err} & \colhead{worst} \\
 & & \colhead{[Gpc]} & & & \colhead{[\%]} & &
\colhead{[\gpcyr]} & & & & & \colhead{[\%]} & \colhead{[\%]} \\
\hline
$0$       & $142$ & $2.28$ &  & $55.0$ & $3.5$  &  & $108.0$ & $2.242$ &  & $6.06$ & $3.86$ & $9.2$  & $39$ \\
$0.9$     & $142$ & $2.23$ &  & $56.9$ & $3.1$  &  & $203.9$ & $2.582$ &  & $4.00$ & $4.64$ & $7.6$  & $28$ \\
$0.99$    & $70$  & $2.10$ &  & $62.7$ & $3.3$  &  & $129.8$ & $2.740$ &  & $4.24$ & $4.36$ & $8.4$  & $26$ \\
$0.999$   & $22$  & $1.88$ &  & $68.4$ & $3.6$  &  & $56.5$  & $2.813$ &  & $5.42$ & $3.88$ & $11.1$ & $22$ \\
$0.9999$  & $22$  & $1.66$ &  & $79.3$ & $3.8$  &  & $21.7$  & $2.847$ &  & $7.33$ & $3.41$ & $4.1$  & $21$ \\
$0.99999$ & $142$ & $1.34$ &  & $127$  & $15.6$ &  & $7.0$   & $2.863$ &  & $9.71$ & $2.67$ & $5.7$  & $31$ \\
\hline
All       & $540$ &        & $1.49$ &  & $4.7$ &  &         &         & $-0.318$ &  &  & $7.6$ & $39$
\end{tabular}
\end{ruledtabular}
\end{table*}

\section{Fits of the maximum distance and sensitivity time-volume}
\label{app:fits}

We provide additional details of the empirical fits of the horizon distance, Eqs.~\eqref{eq:dmax_fit} and~\eqref{eq:F_D_fit}, and the sensitivity time-volume, Eqs.~\eqref{eq:VT_fit} and~\eqref{eq:F_VT_fit}, which we have introduced in Sec.~\ref{ssec:vt_res}, and their use across the \LIMRI's parameter space considered in this paper. 

We remind the reader that the fitting formulae have been derived under the assumption of an observation time $\Tobs=4\,\mathrm{yr}$ and an SNR detection threshold $\rhothr = 15$. $D_L^{(\mathrm{max)}}(\bm\theta)$ and $\VT(\bm\theta)$ do not scale trivially with $\Tobs$ in particular for systems in eccentric orbits. 
While at $z\ll1$ the distance scales trivially with SNR, $\VT(\bm\theta)$ does not: it is an integral over the visibility window, and a change of threshold moves both the horizon and the
duration over which each source is detectable in each redshift shell.

The values of the fitting parameters that enter Eqs.~\eqref{eq:dmax_fit}, \eqref{eq:F_D_fit}, \eqref{eq:VT_fit} and~\eqref{eq:F_VT_fit} are listed in Table~\ref{tab:fits}. 

To compute $D_L^{(\mathrm{max)}}(\bm\theta)$ at values of $e_0$ that are different from those provided in Table~\ref{tab:fits}, one can interpolate $\log D_{L,0}$ and $\log Y_h$ provided in Table~\ref{tab:fits} linearly in $\log(1-e_0)$.

For $\VT(\bm\theta)$, the coefficients $\VT_0$, $Y$ and $v$ are not monotonic in $e_0$ and should not be interpolated.
If one wishes to compute $\VT(\bm\theta)$ at a different value of $e_0$, say $e_0^\prime$, where $e_{0,\mathrm{lo}}<e_0^\prime<e_{0,\mathrm{hi}}$ and $e_{0,\mathrm{lo}}$ and $e_{0,\mathrm{hi}}$ are the bracketing values provided in the table, with corresponding $\VT_{\mathrm{lo}} \equiv \VT(e_{0,\mathrm{lo}})$ and $\VT_\mathrm{hi} \equiv \VT(e_{0,\mathrm{hi}})$, instead of using Eq.~\eqref{eq:VT_fit} directly at $e_0^\prime$ one should interpolate linearly in $\log(1-e_0)$,
\begin{equation}
  \log\VT(\bm\theta) = (1-w)\log\VT_\mathrm{lo} + w\log\VT_\mathrm{hi},
  \label{eq:vt_interp}
\end{equation}
with $w=\log[(1-e_0^\prime)/(1-e_{0,\mathrm{lo}})]/\log[(1-e_{0,\mathrm{hi}})/(1-e_{0,\mathrm{lo}})]$.
We have directly tested the accuracy of Eq.~\eqref{eq:vt_interp} using $36$ additional systems with $e_0\in\{0.3,0.5,0.8,0.95,0.97,0.997\}$ and for six mass pairs spanning $q_3M_4^{5/3}=0.1$--$46$, Eq.~\eqref{eq:vt_interp} has a median error of $2.9\%$ and a maximum of $21\%$, mostly inherited from Eq.~\eqref{eq:VT_fit}.

\bibliographystyle{apsrev4-2-etal}
\bibliography{main}

\end{document}